\documentclass[twocolumn,aps,prb]{revtex4-2}

\usepackage{color}
\usepackage{graphicx}

\begin{document}

\title{Equilibrium current distributions and $W_{\infty}$ gauge theory in quantum Hall systems\\
of conventional electrons and Dirac electrons}
\author{K. Shizuya}
\affiliation{Yukawa Institute for Theoretical Physics\\
Kyoto University,~Kyoto 606-8502,~Japan }

\begin{abstract}
In equilibrium planer systems of Hall electrons, 
such as GaAs heterostructures and graphene, 
support two species of current counterflowing along the system edges, 
as observed recently in experiment using a nanoscale magnetometer.
We examine distinct origins and distinctive features of these equilibrium currents,
with the Coulombic many-body effects taken into account, 
 and derive their real-space distributions. 
 Our basic tool of analysis is a reformulation of quantum Hall systems 
 as a $W_{\infty}$ gauge theory, 
 which allows one to diagonalize the total Hamiltonian according to the resolutions of external probes.
 These equilibrium currents are deeply tied to the orbital magnetization in quantum Hall systems. 
 Special attention is drawn to the case of graphene, especially the neutral $(\nu=0)$ ground state 
 and its intrinsic diamagnetic response that combines with the equilibrium currents 
 to govern the orbital magnetization and its oscillations with filling.   

\end{abstract} 


\maketitle 

\section{Introduction}

Two-dimensional (2D) electron systems such as GaAs heterostructures and graphene 
display fascinating electronic properties 
that attract great attention in both applications and fundamental physics. 
They host the quantum Hall (QH) effect~\cite{PG} in a magnetic field. 
An old and basic subject pertaining to the foundation of the QH effect 
is the presence of edge states and the role they play in the electronic transport in such 2D systems.
In equilibrium the edge states become the only current carriers.
Theoretically, along with explanations of the QH effect 
in the bulk-state and edge-state pictures~\cite{Prange,Laughlin,AA,Halperin},
the current flow in QH samples was actively discussed 
earlier~\cite{MRB,MS,HT,Buttiker,CSG,ThoulessW,AHK, ks_qhe,GV}.
In particular, it was pointed out in a model calculation by Geller and Vignale~\cite{GV} 
that the edge states carry a pair of counterflowing equilibrium currents,
one driven by a local field and another by a density gradient,
that form an alternating pattern along the sample edges. 

Experimentally such local features are not amenable to global transport measurements~\cite{ZTC}.
Scanning probe studies, probing locally the potential profile~\cite{FKHB,WvK} 
or the electron density and dissipation~\cite{LKK,SBHM},
detected the presence of edge states but the current they carry locally remained unidentified for years. 
Only recently it has become possible,  by use of a SQUID-on-tip nanoscale magnetometer
in experiment by Uri {\it et al.}~\cite{UKBL}, to achieve 
direct imaging of the counterflowing equilibrium currents in graphene, 
indeed forming an alternating pattern.

With such early and recent developments in mind, we study, in this paper, 
the equilibrium current distributions in QH systems with gentle edges.  
Our basic tool of analysis is a formulation, as a $W_{\infty}$ gauge theory, 
of QH systems coupled to external probes~\cite{ks_Winf}. 
The $W_{\infty}$ [or $U(\infty)$] transformations mix Landau levels and 
the associated gauge field is expressed as a series of multipoles (or derivatives) 
of external potentials. 
They serve to resolve level mixing according to the resolutions of external probes. 
We use this gauge-theory framework to diagonalize the many-body Hamiltonian 
in such a way that the level spectra and associated currents are directly read from it.
In particular, full use is made of a special $W_{\infty}$ gauge transformation,
that systematizes our analysis in  such a manner that   
electromagnetic gauge invariance is manifest from the start.

We first consider a QH system of conventional 2D electrons (with quadratic dispersion).
Keeping only the lowest multipoles for the current operator
allows one to confirm some key findings of an early analysis of Geller and Vignale, 
i.e., the presence of two species of counterpropagating equilibrium currents along the sample edges.
Those lowest multipoles fix the integrated amount of these currents while it turns out necessary  
to pick up also higher multipoles to derive their real-space distributions. 
Their distinct distributions reveal their distinct origin:
One species, a diamagnetic current flowing fast with a narrow profile, 
derives from quantized cyclotron motion of electrons and 
arises (or survives) only along the periphery of a densely populated domain. 
Another one is essentially a Hall current driven by a local edge potential. 
A close study is also made of the Coulombic many-body effects on the two species of current, 
that clearly reflect their distinctive characters.

Subsequently we examine the case of Dirac electrons in graphene.  
 In graphene the Landau levels are naturally divided into sectors $\{|n|=0,1,2,\cdots\}$, 
with each sector $|n|$ consisting of a pair of electron and hole levels (labelled by $\pm |n|$)
related by  electron-hole ($e$-$h$) symmetry.
The $W_{\infty}$ transformations work to partially diagonalize the Hamiltonian in sectors $\{ |n| \}$.
The real-space current distributions and many-body effects on them in the $|n|\ge 1$ sectors 
show features qualitatively similar to those of conventional electrons
while those in the lowest  ($|n|=0$) Landau level exhibit some peculiar features. 
Special attention is drawn to the neutral $\nu=0$ ground state and 
its intrinsic diamagnetic response of $\lq\lq$relativistic" origin 
that combines with those equilibrium currents to govern the orbital magnetization 
and its oscillations (the de Haas-van Alphene effect) in graphene.

The paper is organized as follows. 
In Sec.~II we review and refine the $W_{\infty}$ gauge-theory formulation of a QH system 
and introduce a special gauge transformation mentioned above. 
In Sec.~III we consider a QH system of conventional electrons 
and clarify some general features of two distinct species of equilibrium current
and their distributions. 
In Sec.~IV we examine the Coulombic many-body effects on them.
In Sec.~V we focus on the case of Dirac electrons in graphene. 
Section VI is devoted to summary and discussion.

\section{Electrons in a Magnetic Field and $W_{\infty}$ gauge theory}

Consider conventional 2D electrons in a magnetic field $B_{z}=B>0$, 
with the  potential $(A_{x}, A_{y}) = (-By,0)$.
The one-body Hamiltonian 
$H= \int dx dy \, \Psi^{\dag} {\cal H}\Psi$, with
\begin{eqnarray} 
{\cal H} &=&  {1\over{2m^{*}}} \{(p_{x}- eBy)^2 
+ p_{y}^2 \} = {1\over{2}}\,  \omega_{c}\, (Y^2 + P^2),
\label{H_two-d_e}
\end{eqnarray}
is essentially a harmonic-oscillator system 
with the normalized coordinate $Y=(y-y_{0})/\ell$ and momentum 
$P= \ell\, p_{y}$ with $[Y,P]= i$, where $\ell \equiv 1/\sqrt{eB}$
is the magnetic length and $y_{0}\equiv \ell^2 p_{x}$.
The electron spectrum forms Landau levels of energy 
$\epsilon_{n}= \omega_{c} (n + {1\over{2}})$ with $\omega_{c}= eB/m^{*}$,
and the eigenmodes 
$\langle x,y|n,y_{0}\rangle = \langle x|y_{0}\rangle\,  \langle y-y_{0}|n\rangle$,
labeled by $n\in (0,1,2,\cdots)$ and  $y_{0}$,
consist of plane waves $\langle x|y_{0}\rangle = e^{ix\,y_{0}/\ell^2}/\sqrt{2\pi \ell^2}$
and the harmonic-oscillator wave functions $\langle y |n\rangle \equiv \phi_{n}(y)$.   
In the $|n,y_{0}\rangle$ basis 
the coordinate ${\bf x}=(x,y)$ is written as~\cite{ks_Winf} 
\begin{eqnarray}
\langle n,y_{0}| x | n', y'_{0} \rangle 
&=& \{ \delta^{nn'}  i\ell^2 \partial/\partial y_{0} + \ell\, P^{nn'} \}\, \delta (y_{0} - y'_{0}),
\nonumber\\
\langle n,y_{0}| y | n', y'_{0} \rangle 
&=& \{  \delta^{nn'} y_{0} + \ell\, Y^{nn'}\}\, \delta (y_{0} - y'_{0}),
\end{eqnarray}
where $(Y, P)$ now stand for numerical matrices in level (or orbital) indices
of the familiar harmonic-oscillator form.

An electron thus undergoes cyclotron motion 
with matrix coordinate  ${\bf X} \equiv (X_{x}, X_{y})= \ell\, (P,Y)$ 
and center motion 
with continuous coordinate ${\bf r}  = (i\ell^2 \partial_{y_{0}}, y_{0})$. 
In what follows we make extensive use of the $|n,y_{0}\rangle$ basis, 
and denote the coordinate ${\bf x}$
as ${\bf x} = {\bf X} + {\bf r}$,
with uncertainty $[X_{x}, X_{y}] =-i\ell^{2}$, $[r_{x}, r_{y}] =i\ell^{2}$ 
and $[X_{i},r_{j}]=0$.

To study the electromagnetic response of the system 
let us introduce weak external potentials $a_{\mu} = (a_{x}, a_{y}, A_{0})$.
They are taken to be slowly varying in space and time, 
and are to be expanded in multipoles (i.e., derivatives) in our analysis. 
A simple yet practically useful choice is to 
suppose that they depend only on one coordinate $y$ (with time $t$ treated implicitly).
They serve to detect a current $j_{x}(y)$ 
driven by an applied local field $E_{y}(y) = -\partial_{y}A_{0}(y) - \partial_{t}a_{y}(y)$.
They also supply a  local magnetic field 
$b_{z}^{({\bf a})} \equiv \partial_{x}a_{y} - \partial_{y}a_{x}\rightarrow -\partial_{y}a_{x}(y)$
normal to a sample.

Passing to the $|n,y_{0}\rangle$ basis via the expansion 
$\Psi ({\bf x}) = \sum_{n, y_{0}} \langle {\bf x}|n,y_{0}\rangle\, \psi^{n}(y_{0})$
yields the Hamiltonian
\begin{eqnarray}
H &=& \int dy_{0}\, \sum_{m,n}\psi^{m \dag}(y_{0})\, {\cal H}^{mn}\, \psi^{n}(y_{0}),
\\
{\cal H} &=& \omega_{c}\, \big\{ (Z^{\dag} -i v^{\dag}) (Z + iv) 
+{\textstyle {1\over{2}}}+ {\textstyle {1\over{2}}}b_{z} \big\} -eA_{0},
\end{eqnarray}
with $Z \equiv (Y+ iP)/\sqrt{2}$, $Z^{\dag} \equiv (Y- iP)/\sqrt{2}$;
$[Z,Z^{\dag}]=1$ and $Z^{mn} \equiv \langle m| Z| n\rangle = \sqrt{n}\, \delta^{m,n-1}$.
Here ${\cal H}$ stands for a matrix ${\cal H}^{mn}\equiv \langle m|{\cal H}|n\rangle$ in orbital labels;
in what follows we adopt such matrix notation 
and frequently suppress summation over repeated level indices.
In ${\cal H}$ we have set $v_{i}= e \ell \, a_{i}$ (or ${\bf v} = e\ell\,  {\bf a})$, 
$v =  (v_{y}+ iv_{x})/\sqrt{2}$, $v^{\dag} =  (v_{y}- iv_{x})/\sqrt{2}$, and
$b_{z} \equiv \ell \nabla\!\times {\bf v} 
=\ell (\partial_{x}v_{y} - \partial_{y}v_{x} )= e\ell^2 b_{z}^{({\bf a)}}$.

Fields $v= v({\bf x})$ and $A_{0}= A_{0}({\bf x})$ are functions of ${\bf x}= {\bf X} + {\bf r}$, 
and are matrices in orbital labels $\{n\}$. 
Let us adopt the Fourier transform to specify their ${\bf x}$ dependence 
and isolate their matrix portion by writing, e.g.,  
\begin{equation}
v({\bf x})= \sum_{\bf p}v_{\bf p} e^{i{\bf p \cdot (X+r)}} 
= e^{i{\bf p \cdot X}}v({\bf r})
\end{equation} 
with $v({\bf r}) \equiv \sum_{\bf p}v_{\bf p} e^{i{\bf p \cdot r}}$; 
$\sum_{\bf p} \equiv \int d^{2}{\bf p}/(2\pi)^{2}$.
In the last line, we regard ${\bf p}$ as a derivative $-i\nabla$ acting on $v({\bf r})$;
$ip_{y}v({\bf r})$, e.g., stands for $\partial_{y}v({\bf x})$, with ${\bf x} \rightarrow {\bf r}$. 
One can rewrite 
$e^{i{\bf p \cdot X}} = e^{i  \ell p Z^{\dag} + i  \ell p^{\dag} Z} =  \gamma_{\bf p} f_{\bf p}$,
with  $ \gamma_{\bf p} =e^{-{1\over{4}}\ell^2 {\bf p}^2}$
and $f_{\bf p} = e^{i\ell pZ^{\dag}} e^{i\ell p^{\dag}Z}$, where
$p = (p_{y} + ip_{x})/\sqrt{2}$ and  $p^{\dag} = (p_{y} - ip_{x})/\sqrt{2}$, 
or $ip \rightarrow (\partial_{y} + i\partial_{x})/\sqrt{2}\equiv \partial$ 
and $ip^{\dag} \rightarrow  \partial^{\dag}$.  
Note, e.g., $[Z, v({\bf x})] = \ell \partial v({\bf x})$ and 
$[Z^{\dag}, v({\bf x})] = -\ell \partial^{\dag} v({\bf x})$.
A function  of ${\bf x}= {\bf X + r}$, e.g., $v({\bf x})$, 
is then expanded in a normal-ordered series of $(Z^{\dag}, Z)$ as
\begin{eqnarray}
v({\bf x})
 &=&\sum_{s=0}^{\infty} D_{s}[\partial]\, 
  \gamma_{\bf p}  (\ell^{2}\partial^{\dag}\partial)^{s}  v({\bf r}),
 \label{vG_expansion}\\
D_{s}[\partial]&=& F_{s} + \sum_{r=1}^{\infty} 
\big\{ F_{s}^{r0} (\ell \partial)^{r}+  F_{s}^{0r}(\ell \partial^{\dag})^{r} \big\}, 
\end{eqnarray}
with 
$F_{s}= (Z^{\dag})^{s}Z^{s}/(s!)^2$, $F_{s}^{r0} =  (Z^{\dag})^{s+r}Z^{r}/\{(s+r)! s!\}$
and $F_{s}^{0r} =(F_{s}^{r0})^{\dag}$;
obviously, $F_{s}$ are diagonal in orbital labels $\{n\}$ while the rest are not;
$ \gamma_{\bf p} = e^{{1\over{4}}\ell^2 \nabla^{2}} = e^{{1\over{2}}\ell^2 \partial^{\dag} \partial}$.
In this way, the matrix portion of $v({\bf x})$ is naturally expanded 
in a series of multipoles of $v({\bf r})$.
The actual values of $\langle m|F_{s}^{r0}|n \rangle =[F_{s}^{r0}]^{mn}$, etc.,  
are readily extracted from the matrix elements
$f^{mn}_{\bf p} = \langle m|e^{i\ell pZ^{\dag}} e^{i\ell p^{\dag}Z}|n \rangle$, 
expressed in terms of the associated Laguerre polynomials, 
\begin{equation}
f^{mn}_{\bf p} = \sqrt{n!/m!}\,  (i\ell p)^{m-n} L^{m-n}_{n}(\ell^{2}p^{\dag}p).
\label{fmn}
\end{equation}
A useful formula is
\begin{equation}
\sum_{s=0}^{\infty}  (-\xi)^{s} [F_{s}^{r0}]^{n+r, n} 
=  \sqrt{n!/(n+r)!}\, L_{n}^{r}(\xi).
\label{FtoL}
\end{equation}

The one-body Hamiltonian ${\cal H}^{mn}$ in $H$ 
is an infinite-dimensional hermitian matrix. 
Let us consider a class of unitary transformations 
$\psi^{n}$ $\rightarrow \psi_{U}^{n} = U^{nm}\psi^{m}$, 
that mix Landau levels $\{ n\}$, with $U$ of the form 
\begin{equation}
U = {\rm exp}\Big[ i \sum_{r,s=0}^{\infty} \alpha_{rs}({\bf r}) (Z^{\dag})^{r}Z^{s}\Big].
\end{equation}  
The basis $\{(Z^{\dag})^{r}Z^{s} \}$ forms the $U(\infty)$ (or $W_{\infty}$) algebra.
It is possible to formulate 2D Hall electron systems as a $W_{\infty}$ gauge theory,
as noted earlier~\cite{ks_Winf}. 
The action 
\begin{equation}
L = 
\int\! dtdy_{0}\, \psi^{\dag} 
\big(  i\partial_{t}  - {\cal H} \big)\, \psi
=\int dt dy_{0}\, \psi_{U}^{\dag} 
\big(  i\partial_{t}  - {\cal H}^{U} \big)\, \psi_{U}
\end{equation}
 is invariant under $W_{\infty}$ transformations 
$\psi \rightarrow \psi_{U} = U \psi$  and  ${\cal H}^{U} = U({\cal H} - i\partial_{t})U^{\dag}$
if the associated \lq\lq $W_{\infty}$ gauge fields" 
$v({\bf x})$ and $A_{0}({\bf x})$ are transformed as
\begin{eqnarray}
v^{U}({\bf r})  &=& U\, v({\bf x}) \, U^{\dag} -iU\, [Z,U^{\dag}],
\nonumber\\
eA_{0}^{U}({\bf r}) &=& U\, \{ eA_{0}({\bf x})  + i \partial_{t}\} U^{\dag}.
\end{eqnarray}
Here only the argument  ${\bf r}$ is retained for $v^{U}$ and $A_{0}^{U}$, 
which are no longer functions of a single ${\bf x} = {\bf r +X}$.

Obviously, $Z + iv$, $Z^{\dag} - iv^{\dag}$ and $\partial_{t} - ie A_{0}$ act as covariant derivatives, 
$U\, (Z + iv)\, U^{\dag} = Z + iv^{U}$, $U\, ( \partial_{t} - ieA_{0} ) U^{\dag} =\partial_{t} - ie A^{U}_{0}$, etc. 
Their commutators 
\begin{eqnarray}
&& [Z + iv({\bf x}), Z^{\dag} - iv^{\dag}({\bf x})] = 1 +  b_{z}({\bf x}),
\nonumber\\
&&[Z + iv({\bf x}),  \partial_{t}-i eA_{0}({\bf x})] =  ie\ell E({\bf x}),
\end{eqnarray}
then reveal the $W_{\infty}$ field strengths $E_{j} = -\partial_{j}A_{0} - \partial_{t} a_{j}$
and $b_{z} =-i \ell (\partial v^{\dag} -\partial^{\dag}v)$, 
which transform covariantly under $U$, $ (E^{U}, b^{U}_{z}) =U (E, b_{z})U^{\dag}$;
$E  \equiv (E_{y} + i E_{x})/\sqrt{2}$.
They coincide with the electromagnetic fields $(E, b_{z})$.

The gauge fields $(v,A_{0})$ transform inhomogeneously under $U$. 
It is intriguing to see what will happen if one eliminates a gauge-variant portion out of $v$. 
See Appendix A for an analysis in this direction. 
The result is a $W_{\infty}$ gauge transformation $G = e^{iS}$ with
\begin{eqnarray}
S &=& \sum_{s=0}^{\infty} \sum_{r=1}^{\infty}\!\! \gamma_{\bf p} (\partial^{\dag}\partial)^{s}
 \big\{ F^{r0}_{s} \partial^{\, r-1} v({\bf r})
+  F^{0r}_{s} (\partial^{\dag})^{r-1} v^{\dag}({\bf r}) \big\} 
\nonumber\\
&& + \sum_{s=0}^{\infty} F_{s+1} 
{\textstyle {1\over{2}} }\, \gamma_{\bf p}\, (\partial^{\dag} \partial)^{s}\nabla\! \cdot\! {\bf v}({\bf r}),
\label{Gfirst}
\\
&=& \textstyle  
 Z^{\dag} v({\bf r}) + Z v^{\dag}({\bf r})
+ {\textstyle{1\over{2}}} Z^{\dag}Z\,  \nabla\! \cdot\! {\bf v}({\bf r})  
\nonumber\\
&&+  {\textstyle{1\over{2}}} Z^{\dag 2}  \partial v({\bf r}) 
+ {\textstyle{1\over{2}}}Z^2 \partial^{\dag} v^{\dag}({\bf r})
+ \cdots. 
 \end{eqnarray}
[For conciseness, we suppress magnetic length $\ell \rightarrow 1$ from now on, 
taking it as a basic length unit, and recover it, when appropriate.]  
Remarkably, the transformed field   $v^{G} \approx v - [Z, S]$, 
to first order in $v$,
is expressed in terms of multipoles of magnetic field $b_{z}({\bf  r})$ alone,
\begin{eqnarray}
v^{G} &\stackrel{O(v)}{=}& 
- i \sum_{s=0}^{\infty} \Big[  {\textstyle {1\over{2}} } F_{s}^{01}
+\sum_{r=1}^{\infty}  F_{s}^{0,r+1}  (\partial^{\dag})^{r}\, \Big]  
\gamma_{\bf p} (\partial^{\dag} \partial)^{s}\, b_{z}({\bf r}), 
\nonumber\\
&=& \textstyle -i   \big \{  {\textstyle {1\over{2}} } Z \, b_{z}({\bf r}) 
 + {1\over{2}}Z^2  \partial^{\dag} b_{z}({\bf r})  + O(\partial^{3})  \big\},
\label{vG} 
\end{eqnarray}
where $O(\partial^{3})$ denotes terms  involving three powers of derivatives or more acting on $v({\bf r})$.
At the same time, $eA_{0}^{G} = eA_{0} + \dot{S} + i[S,eA_{0}] + i{1\over{2}}[S,  \dot{S}] + \cdots$
(with $\dot{S} \equiv \partial_{t}S$), to $O(v)$, 
is expressed in terms of $A_{0}$ and ${\bf E} = (E_{x}, E_{y})$,
 \begin{eqnarray}
 A_{0}^{G} 
 &\stackrel{O(v)}{=}& \gamma_{\bf p} A_{0}({\bf r}) 
 - \sum_{s=0} F_{s+1}\,
{\textstyle {1\over{2}} }  \gamma_{\bf p}\,   (\partial^{\dag} \partial)^{s} \nabla\! \cdot\! {\bf E}({\bf r})
 \nonumber\\
&& - \sum_{s=0} \sum_{r=1} \gamma_{\bf p} (\partial \partial^{\dag})^{s}
 \{F^{r0}_{s} \partial^{\, r-1} E + F^{0r}_{s} (\partial^{\dag})^{r-1} E^{\dag} \},
\nonumber\\
&=& \gamma_{\bf p} A_{0}({\bf r}) 
- Z^{\dag}Z\,  {\textstyle {1\over{2}} }   \nabla\! \cdot\! {\bf E}
- Z^{\dag} E - Z\, E^{\dag}  + \cdots, 
\end{eqnarray}
where $E = E({\bf r})$ and $E^{\dag} = E^{\dag}({\bf r})$.

The gauge transformation $\psi \rightarrow \psi_{G}= G \psi$ leads to the  one-body Hamiltonian 
$H^{G}= \int dy_{0}\psi_{G}^{\dag}{\cal H}^{G}\psi_{G}$, 
with
\begin{eqnarray}
{\cal H}^{G}
&=&  \omega_{c}(Z^{\dag}Z + {\textstyle {1\over{2}}} ) - eA_{0}^{G} + \omega_{c}(v^{G})^{\dag}v^{G}
 \nonumber\\
&&+  \omega_{c}\sum_{s=0}^{\infty}\Big[  (s\!+\!1)D_{s+1}\! + {\textstyle {1\over{2}}} D_{s}\Big]\, 
 (\partial^{\dag}\partial)^s \gamma_{\bf p}\, b_{z}({\bf r}), \ \  \ \ 
\end{eqnarray}
where $D_{s}=D_{s}[\partial]$ for short.
It is clear that ${\cal H}^{G}$ is exactly diagonalized [in orbitals $(m,n)$] to first order in $v$ and $A_{0}$;
the off-diagonal pieces eventually lead to (diagonal) corrections of $O(v^2)$, $O(A_{0}^2)$ and $O(vA_{0})$.
Of our particular interest are $O(vA_{0})$ terms that govern how the current flows 
when the electrons are driven by an electric field $E_{j} = -\partial_{j}A_{0} - \partial_{t} a_{j}$. 
One such $O(vA_{0})$ term  comes from $-eA_{0}^{G} \ni -i[S,eA_{0}+ \dot{S}]$, with the diagonal piece
\begin{equation}
{\cal H}_{vA} = 
\sum_{s=0}^{\infty} F_{s} \{(\partial^{\dag}\partial)^s \gamma_{\bf p}\, {\bf v}({\bf r}) \}\! \times\! e{\bf E}({\bf r}) 
+ O(\nabla^2A_{0});
\end{equation}
${\bf v} \times {\bf E} = v_{x} E_{y}- v_{y} E_{x}$.
Here we have retained only terms involving a single derivative of $A_{0}$, 
assuming its gentle spatial variations.
Accordingly we now try to diagonalize ${\cal H}^{G}$ to $O(v\partial A_{0})$ 
(while keeping full multipoles of $v$).
Let us first eliminate the $O(E)$ off-diagonal piece in ${\cal H}^{G}$,
\begin{equation}
\Delta^{\rm off} {\cal H}^{G} =  e E({\bf r})\, Z^{\dag}+ e E^{\dag}({\bf r})\, Z +O(\partial^2),
\end{equation}
by a further $W_{\infty}$ rotation
$\psi_{G} \rightarrow \hat{\psi} = G_{2}\, \psi_{G} = G_{2} G\, \psi$,
with $G_{2}= e^{iS_{2}}$ and 
\begin{equation}
S_{2}=  -i\beta[b_{z}]\{ E({\bf r}) Z^{\dag} - E^{\dag}({\bf r}) Z\} + \cdots, 
\label{S-two}
\end{equation}
where $\beta[b_{z}] =(e/\omega_{c}) \{ 1- (1 +{1\over{2}} \partial^{\dag}\partial)\,  b_{z}({\bf r})\}$.
The off-diagonal portion in the $\omega_{c} (\cdots)b_{z}({\bf r})$ term of ${\cal H}^{G}$ 
thereby yields another diagonal piece of $O(v\partial A_{0})$,
\begin{eqnarray}
{\cal H}^{(2)}_{vA} &=& -  \,e{\bf E}({\bf r})\cdot \sum_{s=0} \Gamma_{s}  
(\partial^{\dag}\partial)^s \gamma_{\bf p}\nabla\,b_{z}({\bf r}),
\label{H_vA_two}
\end{eqnarray}
 where $\Gamma_{s}= (s+1) F_{s+1} + F_{s}/2$; $\Gamma_{0}=Z^{\dag}Z+ 1/2$.
(Note formulas $[F^{0r}_{s}, Z^{\dag}] = F^{0,r-1}_{s}$, 
$[F^{0r}_{s}, Z] = - F^{0,r+1}_{s-1}$, etc.)

 Let us denote  by $\hat{H} =\int dy_{0} \hat{\psi} \hat{\cal H}\hat{\psi}$ 
 with $\hat{\cal H} = G_{2}{\cal H}^{G}G_{2}^{-1}|^{\rm diag}$ 
 the resulting Hamiltonian diagonal to $O(v)$, $O(A_{0})$ and $O(v \nabla A_{0})$.
For static potentials (with $\dot{v} = \dot{A}_{0}=0$), 
$\hat{\cal H}$ is neatly written as 
\begin{eqnarray}
\hat{\cal H} &=&  \omega_{c}(Z^{\dag}Z + {\textstyle {1\over{2}}} )  -e \hat{F}A_{0}  
+  \omega_{c} \hat{\Gamma}\, b_{z}({\bf r})
 \nonumber\\
&& - e {\bf E}({\bf r})\! \times \hat{F}\, {\bf v}({\bf r})  
- e   {\bf E}({\bf r}) \cdot \hat{\Gamma}\, \nabla\,b_{z}({\bf r}),
\label{hatH}
\end{eqnarray}
with 
$\hat{F} = \sum_{s=0}^{\infty}F_{s} (\partial^{\dag}\partial)^s \gamma_{\bf p}$
 and $\hat{\Gamma} = \sum_{s=0}^{\infty}\Gamma_{s} (\partial^{\dag}\partial)^s \gamma_{\bf p}$.
Noting Eq.~(\ref{FtoL}) one can project $\hat{F}$ and $\hat{\Gamma}$ to each level $n$,
\begin{eqnarray}
k_{n}(\xi) &\equiv& [\hat{F}]^{nn} =  e^{- {1\over{2}} \xi}\, L_{n}(\xi), 
\nonumber\\
h_{n}(\xi) &\equiv& {[\hat{\Gamma}]^{nn}} 
= e^{- {1\over{2}} \xi}\, \{L_{n-1}^{1}(\xi) + {\textstyle {1\over{2}}}\,  L_{n}(\xi) \},
\end{eqnarray}
with $\xi \equiv -\ell^2\partial^{\dag}\partial$ and $L^{1}_{n-1}(\xi) =  -  (d/d\xi) L_{n}(\xi)$;
$k_{n}(\xi)=1 - (n+ {1\over{2}}) \xi + {1\over{8}} (2n^{2} + 2n +1) \xi^2+ \cdots$ 
and $h_{n}(0)=n + {1\over{2}}$.
Note that $h_{n}(\xi) = - (d/d\xi) k_{n}(\xi)$ holds.

\section{Edge currents}

 \begin{figure}[tb]
\begin{center}
\includegraphics[scale=.7]{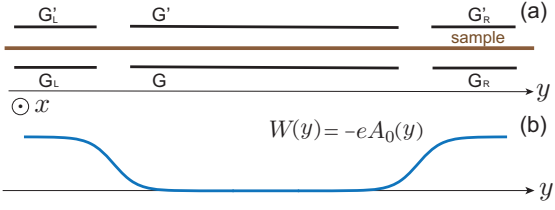}
\end{center}
\vskip-.7cm
\caption{  
(a) Cross section of a sample under the control of gates. 
(b)~Potential wall induced by a static  potential $A_{0}(y)$.
}
\end{figure}

In this section we study equilibrium current distributions in a QH system with edges.
Figure~1 illustrates the Hall bar sample we consider.
It extends homogeneously in the $x$ direction while, in the $y$ direction, 
it is divided into three domains under the control of three sets of gates (G, G').
In each domain the potential $A_{0}(y)$ is taken to be flat, 
except for the (left/right) edge portions 
where $A_{0}(y)$  connects the adjacent domains smoothly.
We thus use it to simulate a potential \lq\lq wall" $W(y) \equiv -e A_{0}(y)$ 
that confines electrons in some lower Landau levels. 
We suppose a gentle edge $\sim \partial_{y}A_{0}(y)$ so that the basic features 
of bulk Landau levels remain intact.
In such a static setting it suffices to adopt a potential $v_{x}(y) = e\ell a_{x}(y)$, 
that depends only on $y$, to detect the ($x$-averaged) current 
$j_{x}(y)= (1/L_{x})\int dx\,  j_{x}(x,y)$ (with $L_{x}= \int dx$) 
flowing along the gate-induced edges.

The electric current ${\bf j} =(j_{x}, j_{y}) = - \delta H/\delta {\bf a}$ 
is read from the Hamiltonian through terms linear in $a_{j}$.
The current  and charge, coupled to potentials $(v,A_{0})$,  
in the original $\psi^{n} (y_{0})$ basis induce Landau-level mixing. 
The transformation $\psi \rightarrow \hat{\psi} = G_{2}G\, \psi$ 
resolves such level mixing to $O(v)$ and $O(v \partial A_{0})$ 
for $\hat{\cal H}$ in Eq.~(\ref{hatH}), 
and the current distribution is directly read from it 
by taking a ground-state expectation value $\langle \hat{H} \rangle$.

 Let us now take a static setting ${\bf r} \rightarrow y_{0}$, 
 ${\bf v}({\bf r}) \rightarrow v_{x}(y_{0})$ and 
 $b_{z} \rightarrow - \ell\partial_{y_{0}}v_{x}(y_{0}) \equiv - \ell\, v'_{x}(y_{0})$ and   
 project $\hat{\cal H}$ to each level $n$.  Then $\hat{H}$ is cast in the form
\begin{eqnarray}
\hat{H} &=&
 \int dy_{0} \sum_{n} \{ \epsilon_{n}(y_{0}) +  {\cal H}^{v}_{n}(y_{0})\}\, \hat{\rho}_{n}(y_{0}),
\label{hatH_cap}
\\
\epsilon_{n}(y_{0}) &=&  (n  +  {\textstyle {1\over{2}}} ) \, \omega_{c}  + k_{n}(\xi)W(y_{0}), 
\\
&\approx &
  (n  +  {\textstyle {1\over{2}}} )  
\{\omega_{c}  +  {\textstyle {1\over{2}}} \ell^2  W''(y_{0})\} + W(y_{0}), 
\nonumber\\
{\cal H}^{v}_{n}(y_{0})&=&
-  \omega_{c}\ell\, h_{n}(\xi)\, v'_{x}(y_{0})
\nonumber\\  &&
+  \ell W'(y_{0})\,\{k_{n}(\xi) -2\xi\, h_{n}(\xi)\}\, v_{x}(y_{0}),\ \ \ \ 
\label{Hvj}
\end{eqnarray}
where $\xi \equiv -\ell^2\partial^{\dag}\partial={1\over{2}} \ell^{2} p_{y}^{2}$ 
is a derivative operator acting on $v_{x}(y_{0})$ and $W(y_{0})$;
$W'(y_{0})\equiv \partial_{y_{0}}W(y_{0})$.
Here $\hat{\rho}_{n}(y_{0}) = \hat{\psi}^{n \dag}(y_{0}) \hat{\psi}^{n}(y_{0})$ 
is the electron density of the $(n, y_{0})$ mode 
with the spectrum $\epsilon_{n}(y_{0})$ in the presence of a potential wall $W(y)$.

Varying $\hat{H}$ with respect to $a_{x}(y_{0})$ yields 
the $x$-averaged current density in the eigenmode space $\{n, y_{0}\}$. 
Let us, for the moment, take only the lowest multipoles, 
\begin{eqnarray}
j^{(0)}_{x}[y_{0}] &=& -(e\ell^{2}/L_{x})
\sum_{n} I_{n}^{(0)}(y_{0}),
\nonumber\\
I_{n}^{(0)}(y_{0}) &=& 
\omega_{c} (n+ {\textstyle {1\over{2}}) }\,\partial_{y_{0}}  \hat{\rho}_{n}(y_{0}) 
+ W'(y_{0}) \hat{\rho}_{n}(y_{0}).
\label{jx_zero}
\end{eqnarray}
Actually, $j_{x}^{(0)}[y_{0}]$, in this operator form, 
confirms the result of an earlier Green-function analysis of Ref.~\cite{GV}. 
The current (density) consists of two components,\break
(i)~an $\lq\lq$edge" current  $j^{\rm (c)} \propto \partial_{y_{0}}  \hat{\rho}_{n}(y_{0})$ 
driven by a change in the electron density
and  (ii)~a $\lq\lq$bulk" current $j^{\rm (d)} \propto W'$
driven by a field $W'(y_{0}) = e E_{y}(y_{0})$.
For clarity, we call $j^{\rm (c)}$ a $\lq\lq$circulating" current 
since it comes from the cyclotron motion of electrons, 
as elaborated on later. 
Similarly, we call $j^{\rm (d)}$ a $\lq\lq$drift" current.
The current carried by a given ground state is calculated 
by taking an expectation value $\langle j_{x}[y_{0}] \rangle$.
Obviously $\langle j^{\rm (d)}[y_{0}] \rangle$ and $\langle j^{\rm (c)}[y_{0}] \rangle$
both vanish deep in the sample interior,
where $W'(y_{0})\rightarrow 0$ and $\langle \hat{\rho} (y_{0})\rangle \rightarrow $ constant.

For clarity, let us hereafter focus on one edge of a sample. 
We suppose that $W(y) =0$ deep in the \lq\lq bulk" $y \ll 0$ and 
that $W(y)$  rises as  $y \rightarrow 0$ and accommodates 
a few lower Landau levels  in the edge region $y \sim 0$ and inward.
Each filled level is characterized by a filled domain $\{y_{0}; y_{0}  \le y_{0;n}^{+} \}$ 
and the boundary $y_{0;n}^{+}$ is fixed from the spectrum 
$\epsilon_{n}(y_{0;n}^{+}) = \epsilon_{\rm F}$
for a given value of the Fermi energy $\epsilon_{\rm F}$. 
Each filled domain has a constant density 
$\langle \hat{\rho}_{n}(y_{0}) \rangle/L_{x} = \bar{\rho} =1/(2\pi \ell^2)$ 
owing to Fermi statistics.

Due to this simple density profile in the  $y_{0}$ space,
the total amount of current $j_{x}[y_{0}]$ per edge $y \sim 0$ 
is  calculable from the lowest multipole $j_{x}^{(0)}[y_{0}]$.
For each filled level $n$, 
\begin{eqnarray}
J^{\rm (c)}_{n} = \int dy_{0}\, \langle  j_{n}^{\rm (c)}[y_{0}] \rangle 
&=&  {e \omega_{c}\over{2\pi}} (n+ 1/2) ,
\label{amount_edgeC}
\\
J^{\rm (d)}_{n} =\int dy_{0}\, \langle  j_{n}^{\rm (d)}[y_{0}] \rangle 
&=& - {e\over{2\pi}}\,W(y_{0;n}^{+}).
\label{amount_bulkC}
\end{eqnarray}
The drift current $J^{\rm (d)}= \sum_{n}J^{\rm (d)}_{n}$ is governed 
by Hall voltages $\propto W(y_{0;n}^{+})$ acting on each level $n$ 
across the edge region and by Hall conductance 
$\sigma_{xy} = -e^2/(2\pi \hbar)$ per level.
The two currents in general flow in opposite directions, 
$J^{\rm (c)}_{n} > 0$ and $J^{\rm (d)}_{n} <0$ at the present edge.  
In equilibrium, they simply circulate along the sample edges.

It is the current distribution $\langle  j_{x}(y) \rangle$ in the real space ${\bf x}$ that is directly observable.
To derive it  let us go back to $\hat{H}$ in Eq.~(\ref{hatH_cap}) and try to switch from $v_{x}(y_{0})$ 
to the potential $v_{x}(y)$ in the real $y$ space  through its Fourier transform $v_{x}[p_{y}]$.
One can rewrite, e.g., 
\begin{equation}
k_{n}(\xi)\, v_{x}(y_{0}) 
=  \int dy\,  v_{x}(y)  \sum_{q} e^{iq (y_{0}-y)} k_{n}( {\textstyle {1\over{2}}} \ell^{2} q^2);
\end{equation}
$\sum_{q} = \int dq/(2\pi)$.
It turns out that the Fourier transforms of $k_{n}(\xi)$ and $h_{n}(\xi)$ 
are related to the harmonic-oscillator wave functions
$\phi_{n}(y) = e^{-{1\over{2}} (y/\ell)^2}\, H_{n}(y/\ell)/\sqrt{n!\, 2^{n}\sqrt{\pi}\,  \ell}$,
\begin{eqnarray}
&&\sum_{q}e^{-i q y}\big\{ k_{n}({\textstyle{1\over{2}}} q^2),  h_{n}({\textstyle{1\over{2}}} q^2) \big\}
=\big\{ |\phi_{n}(y)|^2, \Lambda_{n}(y) \big\}, \ 
\label{formula_one}
\\
&&\Lambda_{n}(y) = {\textstyle {1\over{2}} } |\phi_{n}(y)|^2 +  |\phi_{n-1}(y)|^2 +\cdots +   |\phi_{0}(y)|^2,
\\
&&\sum_{q}e^{-i q y -{1\over{4}} q^2} L_{n-1}^{1}({\textstyle{1\over{2}}} q^2) 
= \sum_{m=0}^{n-1} |\phi_{m}(y)|^2. 
\label{formula_dLn}
\end{eqnarray}
See Appendix B for a derivation of these formulas.
From the equality $\partial_{\xi}k_{n} (\xi) = - h_{n}(\xi)$ follows the relation
\begin{equation}
\ell^2\partial_{y}\Lambda_{n}(y) = -y  |\phi_{n}(y)|^2,
\label{formula_two}
\end{equation}
which then leads to the Fourier transform
 $R_{n}(y) =\sum_{q}e^{-i q y}
 \big\{ k_{n}({\textstyle{1\over{2}}} q^2) - q^2  h_{n}({\textstyle{1\over{2}}} q^2) \big\}$,
with 
\begin{equation}
R_{n}(y) =  |\phi_{n}(y)|^2 -\partial_{y}\{ y |\phi_{n}(y)|^2\}. 
\label{formula_R}
\end{equation}
Those fields $\{ \phi_{n}(y)\}$ are localized in $y$ with a spread of a few magnetic lengths.

The real-space current operator is thereby written as
\begin{eqnarray} 
&&j_{x}(y) =-{e\ell^{2}\over{L_{x}}}\sum_{n}\int dy_{0}\, \hat{\rho}_{n}(y_{0})  \hat{I}_{n} (y-y_{0}),
\nonumber\\
&&\hat{I}_{n}(y) = \omega_{c}\partial_{y}\Lambda_{n}(y)  + W'(y_{0})\, R_{n}(y).
\end{eqnarray}
This leads to the current distributions of each filled  level $n$
 in the edge region $y \sim 0$,
\begin{eqnarray}
\langle j_{n}^{\rm (c)}(y) \rangle 
&=& {e\omega_{c}\over{2\pi \ell^2}} \int^{y_{0;n}^{+}}\! dy_{0} (y-y_{0}) |\phi_{n}(y-y_{0})|^2,  
\label{jeyQH}
\\
 \langle j_{n}^{\rm (d)}(y)  \rangle 
 &=& -{e\over{2 \pi}} \int^{y_{0;n}^{+}}\! dy_{0}\, W'(y_{0})\,  R_{n}(y-y_{0}).\ \ \ 
\label{J_bulk}
\end{eqnarray}
In view of Eq.~(\ref{formula_two}), the distribution of the circulating current
$j^{\rm (c)}(y) =\sum_{n} j_{n}^{\rm (c)}(y)$ 
is explicitly evaluated,
\begin{equation}
\langle j^{\rm (c)}(y) \rangle 
= {e \omega_{c}\over{2\pi}}\,\sum_{n} \Lambda_{n}(y- y_{0;n}^{+}), 
\label{je_local}
\end{equation}
with $\int dy\,\langle j^{\rm (c)}(y) \rangle
=  (e \omega_{c}/2\pi) \sum_{n}(n+1/2)$.
Its profile is localized only in the vicinity of the boundary positions $y = y_{0;n}^{+}$ 
of filled levels with a spread of a few magnetic lengths
and each pronounced profile of height $\propto (n+ 1/2)$ moves towards the edge 
with increasing filling $\sim \epsilon_{\rm F}$.

This  $j^{\rm (c)}(y)$ comes from the $- \partial_{y_{0}} v_{x}(y_{0}) \sim  b_{z}(y_{0})$ term
in ${\cal H}^{v}_{n}(y_{0})$ of Eq.~(\ref{Hvj}), 
which represents a magnetic moment 
\begin{equation}
m_{n} = -e\ell^2\omega_{c}\, h_{n}(\xi\rightarrow 0) = -e\ell^2 \omega_{c}(n+ 1/2) < 0
\end{equation}
induced by an orbiting electron of level $n$.
The associated circulating current cancels out locally in a filled domain 
[as seen from Eq.~(\ref{jeyQH})] 
while it survives as $j^{\rm (c)}(y)$ along the periphery.
Actually it is instructive to see this by extracting the magnetization density
from  ${\cal H}^{v}_{n}(y_{0})$,
\begin{equation}
 M_{n}^{z}(y) =  - (e \ell^2\omega_{c}/L_{x})  \int dy_{0}\,  \hat{\rho}_{n}(y_{0})\, \Lambda_{n}(y- y_{0}).
\end{equation}
The current ${\bf j}^{({\rm m})}= \nabla \times {\bf M}$ associated~\cite{Hirst} 
with magnetization ${\bf M}$ then reads $j^{\rm (m)}_{x}(y) = \sum_{n}\partial_{y}M_{n}^{z}(y)$,
which precisely yields $j^{\rm (c)}(y)$;  $j^{\rm (c)}(y)$ is diamagnetic in nature~\cite{Peierls} 
with $m_{n}<0$.
It is clear now why $j_{n}^{\rm (c)}(y)$ is localized only in the periphery of a densely populated domain 
with a small spread and carries a fixed amount 
$J^{\rm (c)}_{n} =(e \omega_{c}/2\pi) (n + 1/2) = -\bar{\rho}\,  m_{n}$ per level;
these properties come from the Landau quantization of cyclotron motion. 
Note that $J^{\rm (c)}_{n}$ is directly related to the magnetization (in the bulk)
$M_{n}^{\rm (c)}\equiv \langle M_{n}^{z}(y) \rangle = -J_{n}^{\rm (c)}$.

As for the drift current $\langle j_{n}^{\rm (d)}(y)  \rangle$, let us first note that, 
in its $y_{0}$ integral, the $-\partial_{y}\{ y |\phi_{n}(y)|^2\}$ portion of $R_{n}(y)$ is 
explicitly integrated [to $O(W')$] to  yield 
\begin{equation}
-W' (y_{0;n}^{+})\,   (y-y_{0;n}^{+})\, |\phi_{n}(y-y_{0;n}^{+})|^2,
\label{jb_integrable}
\end{equation}
which is sizable and {\it oscillating} only in the vicinity of $y =y_{0;n}^{+}$ 
and supports {\it no} net current. 
It is further seen that $R_{n}(y)$ differ from $R_{0}(y)$, 
or equally, $k_{n}({1\over{2}}q^2)$ differ from  $k_{0}({1\over{2}}q^2)= e^{-{1\over{4}}q^2}$, 
by such integrable components $(q^2)^{m}e^{-{1\over{4}}q^2}$.
Accordingly, the drift current $\langle j_{n}^{\rm (d)}(y) \rangle$ of each filled level 
shows a {\it gradual} and {\it universal} growth
\begin{equation}
\langle j_{n}^{\rm (d)}(y)  \rangle 
= -(e/2 \pi)  W'(y) + O(W''') 
\label{jby_local}
\end{equation}
toward the edge 
and goes to zero rapidly around the boundary $y \sim y_{0;n}^{+}$. 
Such a localized level-specific rapid change, in practice, is hardly visible 
because it overlaps with a prominent profile $\Lambda_{n}(y- y_{0;n}^{+})$ 
of $j_{n}^{\rm (c)}(y)$.

Actually, it is somewhat arbitrary how to divide the current $j_{x}(y)$ 
into  $j^{\rm (c)}(y)$ and $j^{\rm (d)}(y)$.  
One may include, e.g.,  the integrable portion in Eq.~(\ref{jb_integrable}) 
into $j_{n}^{\rm (c)}(y)$, 
without affecting the total amount $J^{\rm (c)}_{n}$.

The drift current $j_{n}^{\rm (d)}(y)$  varies in distribution depending 
on the shape of edge potential $W(y)$. 
Still its integrated amount $J_{n}^{\rm (d)} = \int dy\, \langle j_{n}^{\rm (d)}(y) \rangle$ 
is essentially fixed by the filling at the edge $\sim \epsilon_{\rm F}$,
\begin{equation}
J_{n}^{\rm (d)} =  - (e/2\pi)\,   W(y_{0;n}^{+})
\approx  - (e/2\pi)\, (\epsilon_{\rm F} -  \epsilon_{n})
\label{Jd_Ef}
\end{equation}
for $\epsilon_{\rm F} >  \epsilon_{n}= (n+1/2)\, \omega_{c}$. 
In contrast, the circulating current 
$J_{n}^{\rm (c)} = (e\omega_{c}/2\pi)\, (n + 1/2)$ is insensitive to $\epsilon_{\rm F}$, 
and easily evades detection in global measurements.

It will be worth noting here that Eq.~(\ref{jeyQH}) follows from 
the ($x$-integrated) current in the original $\Psi({\bf x})$ basis 
\begin{equation}
\int dx\,j_{x}({\bf x}) =- e\omega_{c}\int dx\,  \Psi^{\dag}({\bf x}) (y_{0} -y)\Psi({\bf x})
\end{equation}
by substituting an eigenmode 
$\Psi_{N}({\bf x}) \approx  \langle x|y_{0}\rangle  \phi_{n}(y-y_{0})$ valid to $O(E^{0})$,
constructed from the $N  \equiv (n,y_{0})$ mode of $\hat{\cal H}$ via 
$\Psi({\bf x}) =\sum_{N'} \langle {\bf x}|N'\rangle [(GG_{2})^{-1}]^{N'N}\hat{\psi}_{N}$.
It is clear that the step $\hat{\psi}^{n}(y_{0}) \rightarrow \Psi_{N}({\bf x})$ 
offers an alternative path to the current distributions 
in Eqs.~(\ref{jeyQH}) and~(\ref{J_bulk}).

For numerical simulations we adopt a potential of the form 
\begin{equation}
W(y) = W_{\rm h}\, \{1 + {\rm tanh}(\lambda\,  y/\ell) \}/2
\label{PotWall}
\end{equation}
and take  $W_{\rm h}= 6\,  \omega_{c}$ and $\lambda=1/20$;
$\ell W'(0)/ \omega_{c} \sim  0.15$ and $W''(y)$ is negligibly small.
As seen from Fig.~2(a), 
this potential accommodates the edge modes of the $n=(0,1,2)$ levels 
over the domain $-50\ell  \lesssim y  \lesssim 0$ of the $y$ axis. 
We write $\epsilon_{\rm F} = \omega_{c} (n_{\rm f} + 1/2)$ and use an effective factor 
$n_{\rm f}$ to specify their filling at the edge.

 \begin{figure}[tbp]
\begin{center}
\includegraphics[scale=.54]{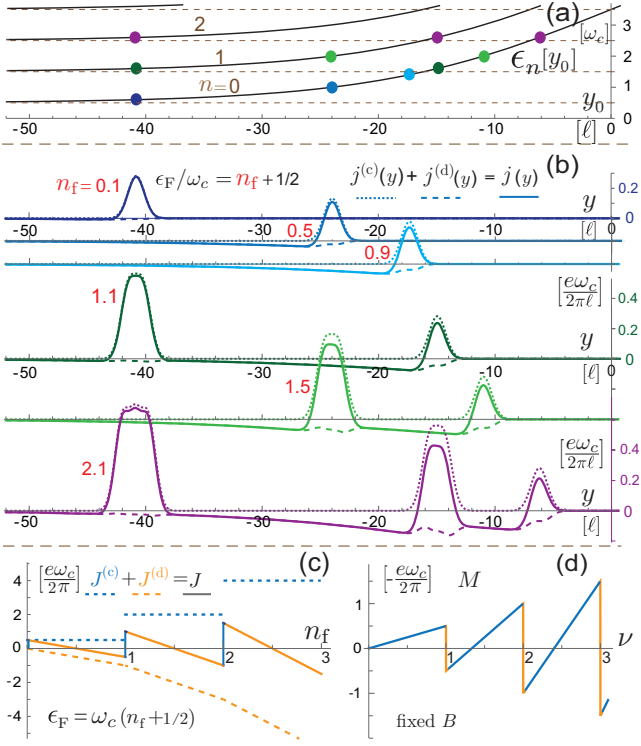}
\end{center}
\caption{(a) Spectra of edge modes. (b)~Current distributions, 
$j^{\rm (c)}(y) + j^{\rm (d)}(y)$, at some edge fillings 
$n_{\rm f}  \in  [0.1, 2.1]$. 
The drift current $j_{n}^{\rm (d)}(y)$ gradually grows $\propto W'(y)$ 
toward the edge boundary $y\sim y_{0;n}^{+}$.
The circulating one $j^{\rm (c)}(y)$ keeps level-specific localized profiles $\Lambda_{n}(y-y_{0;n}^{+})$,
which move with $y_{0;n}^{+}$ as $n_{\rm f}$ is varied.
(c)~$j^{\rm (c)}(y)$ and $j^{\rm (d)}(y)$ compete in the total amount
with increasing filling $n_{\rm f}$. 
(d)~Orbital magnetization $M$ oscillates as a function of the total filling factor $\nu$. }
\end{figure}

Figure~2(b) shows the equilibrium current distributions 
at some fillings $n_{\rm f} \in [0.1, 2.1]$.
The diamagnetic circulating currents $\langle j_{n}^{\rm (c)}(y) \rangle$ 
always flow fast with sharp profiles $\propto \Lambda_{n}(y- y_{0;n}^{+})$
along the periphery of filled domains and 
their positions are in one-to-one correspondence with the spectra [in 2(a)].
The drift currents $\langle j_{n}^{\rm (d)}(y) \rangle$, 
acting as paramagnetic ones, arise only in the edge region,
and gradually grow with a broad profile $\propto -W'(y)$ 
toward the edge boundaries $y \sim y_{0;n}^{+}$.
In the edge region the circulating currents $j_{n}^{\rm (c)}(y)$ appear one after another 
at integer intervals $\Delta n_{\rm f}=1$ 
over a broad background of $j^{({\rm d})}(y)$ of opposite polarity,
thus forming an alternating pattern of current channels,
and they move toward the edge with increasing filling $n_{\rm f}$,
in qualitative agreement with observations~\cite{UKBL}.
While $j^{\rm (c)}$ and $j^{\rm (d)}$ locally differ in distribution, 
they compete in the total amount, as seen from Fig.~2(c); 
$J = J^{\rm (c)} +J^{\rm (d)}$ changes sign across
$n_{\rm f} \approx 0.5, 1.5, 2.5, \cdots$.
In terms of the total filling factor $\nu$,
$J_{n}^{\rm (c)}$ begins to flow for $\nu > n$ 
while $J^{\rm (d)}_{n}$ arises slightly below $\nu =n+1$
and increases with $\nu$.  
Figure~2(d) depicts how the associated magnetization 
$M = M^{\rm (c)} + M^{\rm (d)}= -J^{\rm (c)} - J^{\rm (d)}$
(in units of $-e\omega_{c}/2\pi$) oscillates with increasing $\nu$.

\section{Coulomb interaction}

In this section we study many-body effects on current distributions.
The Coulomb interaction is denoted as
\begin{equation}
V_{c}[\rho] = {1\over{2}} \sum_{\bf p} v^{\rm C}_{\bf p} :\rho_{\bf -p}\, \rho_{\bf p}: \ ,
\end{equation}
with the potential $v^{\rm C}_{\bf p}= 2\pi \alpha_{e}/(\epsilon_{\rm b} |{\bf p}|)$,
$\alpha_{e} \equiv e^{2}/(4 \pi \epsilon_{0})$ and 
the substrate dielectric constant $\epsilon_{\rm b}$;
normal ordering stands for 
${:\! (\psi^{m \dag} \psi^{n}) (\psi^{m'\dag}\psi^{n'})\! :} \, 
\sim \psi^{m'\dag} \psi^{m\dag} \psi^{n} \psi^{n'}$.
The electron density 
$\rho_{-{\bf p}} =\int d^{2}{\bf x}\,  e^{i {\bf p\cdot x}}\, \Psi^{\dag} \Psi$ 
is rewritten as
\begin{eqnarray}
\rho_{-{\bf p}} &=&
 \int dy_{0}\,  [e^{i{\bf p\cdot X}}]^{mn}\, {\cal R}^{mn}_{\bf -p}(y_{0}),
\\
{\cal R}^{mn}_{\bf -p}(y_{0}) &=& \psi^{m\dag}(y_{0})\,e^{i{\bf p}\cdot {\bf r}}\psi^{n}(y_{0}),
\label{U_p}
\end{eqnarray}
where $[e^{i{\bf p\cdot X}}]^{mn} = \gamma_{\bf p}f^{mn}_{\bf p}$ 
with $f^{mn}_{\bf p}$ defined in Eq.~(\ref{fmn}).
The charge operators ${\cal R}^{mn}_{ -{\bf p}}$ and form-factor matrices $[e^{i{\bf p\cdot X}}]^{mn}$
both obey the $W_{\infty}$ algebra~\cite{GMP}.

Upon the $W_{\infty}$ transformation $\psi \rightarrow \hat{\psi} \equiv G_{2}G\psi$, 
the charge density $\rho_{-{\bf p}} = \hat{\rho}_{-{\bf p}} + \delta \hat{\rho}_{-{\bf p}}$
acquires modifications of $O(v)$, $O(vA'_{0})$, etc. 
Thus, in the $\hat{\psi}$ basis, the current operator directly depends on the Coulomb interaction.

Let us examine such modifications in a static setting of potentials 
${\bf v}({\bf r}) \rightarrow v_{x}(y_{0})$ and $A_{0}({\bf r}) \rightarrow A_{0}(y_{0})$.
For simplification, we retain terms up to $O(v'_{x})$ and $O(A'_{0})$ 
for $ \delta \hat{\rho}_{-{\bf p}}$, 
and thus consider [as done for $j_{x}^{(0)}[y_{0}]$ in Eq,~(\ref{jx_zero})] 
the amount of edge currents rather than their detailed spatial distributions.
The relevant correction $\delta \hat{\rho}_{-{\bf p}}$ consists of three terms,
$\delta_{1} \hat{\rho}_{-{\bf p}} + \delta_{2} \hat{\rho}_{-{\bf p}} 
+ \delta_{2} \delta_{1}\hat{\rho}_{-{\bf p}}$:
(i)~$\delta_{1} \rho_{-{\bf p}}= i \int dy_{0}\, \hat{\psi}^{\dag}[S,  e^{i{\bf p}\cdot {\bf x}}]\, \hat{\psi}$ 
comes from the first rotation $G=e^{iS}$ of Eq.~(\ref{Gfirst}),
\begin{eqnarray}
\delta_{1} \hat{\rho}_{\bf -p} &\approx& 
 i \int \! dy_{0}\,  \hat{\psi}^{\dag}\,\Xi^{(1)}_{\bf p} e^{i{\bf p\cdot X}}  e^{i{\bf p\cdot r}} \hat{\psi} 
 \sim O(v),
\\
 \Xi^{(1)}_{\bf p} &=& p_{y} \{v_{x}(y_{0})   - { \textstyle{1\over{2}}} p_{x} v'_{x}(y_{0}) \} 
+  {\textstyle {1\over{2}}} v'_{x}(y_{0})\,  ({\bf p} \cdot\! {\bf X}).\ \ \
\end{eqnarray}
(ii)~$\delta_{2} \rho_{-{\bf p}}= i \int dy_{0}\, \hat{\psi}^{\dag}[S_{2},  e^{i{\bf p}\cdot {\bf x}}]\, \hat{\psi}$
comes from the second rotation $G_{2}$  with $S_{2} \approx C(y_{0})\, P$ of Eq.~(\ref{S-two}),
\begin{eqnarray}
\delta_{2}\hat{\rho}_{\bf -p} &\approx& 
i \int \! dy_{0}\,  \hat{\psi}^{\dag} \,\Xi^{(2)}_{\bf p} e^{i{\bf p\cdot X}}  e^{i{\bf p\cdot r}} \hat{\psi}
\sim O(A'_{0}),
\\
\Xi^{(2)}_{\bf p} &=& [S_{2}, e^{i{\bf p \cdot x}}]e^{-i{\bf p \cdot x}}= p_{y} C'(y_{0}),
\label{Xi-two}
\end{eqnarray}
where $C(y_{0}) =(e/\omega_{c}) \{1-b_{z}(y_{0})\} A_{0}(y_{0})$ and  
$C'= \partial_{y_{0}}C$.\break
(iii)~$\delta_{2}\delta_{1}\hat{\rho}_{-{\bf p}} =i^2 \int \! dy_{0}\,  
\hat{\psi}^{\dag}\,[S_{2}, \Xi^{(1)}_{\bf p}\, e^{i{\bf p\cdot x}}] \,  \hat{\psi}$ 
denotes corrections of $O(vA'_{0})$ obtained from $\delta_{1} \hat{\rho}_{\bf -p}$ 
by a further rotation $G_{2}$,
\begin{equation}
\delta_{2}\delta_{1} \hat{\rho}_{\bf -p} \approx  - \int \! dy_{0}\,  
\hat{\psi}^{\dag}\,\{ [S_{2}, \Xi^{(1)}_{\bf p}] + \Xi^{(1)}_{\bf p}\, \Xi^{(2)}_{\bf p}\}\,
e^{i{\bf p\cdot x}}\,  \hat{\psi}, 
\label{dtwo Vc}
\end{equation}
where $ [S_{2}, \Xi^{(1)}_{\bf p}] \approx -i p_{y}{1\over{2}}v'_{x}(y_{0}) C'(y_{0})$.
The resulting modifications to the interaction 
$V_{c}[\rho] = V_{c}[\hat{\rho}] + \delta_{\rm I} V_{c} + \delta_{\rm II} V_{c}$
are divided into the following two sets, 
\begin{eqnarray}
\delta_{\rm I} V_{c} &=&
\sum_{\bf p}v^{\rm C}_{\bf p}  :\!\big \{ (\delta_{1} \hat{\rho}_{\bf -p}) \, \hat{\rho}_{\bf p}
+ (\delta_{2} \hat{\rho}_{\bf -p}) \, \hat{\rho}_{\bf p}\big\}\!:,
\nonumber\\
\delta_{\rm II} V_{c} &=&
\sum_{\bf p}v^{\rm C}_{\bf p} :\! \big \{ (\delta_{2} \hat{\rho}_{\bf -p}) \, \delta_{1}\hat{\rho}_{\bf p}
+ \hat{\rho}_{\bf -p}\, \delta_{2}\delta_{1} \hat{\rho}_{\bf p} \big\}\! :.
\label{deltaV}
\end{eqnarray}
Here $\delta_{\rm II} V_{c} \propto v_{x} A'_{0}$ represents two different sources of 
the drift current $j^{\rm (d)}$ driven by field $E_{y} = - A'_{0}(y)$.
They combine to  essentially vanish, as we see below.

In general, $V_{c}[\rho]$ has both direct and exchange interactions at the quantum level.
The direct interaction plays no role when a neutralizing background is taken into account. 
For 2D electrons in a magnetic field, it is possible 
to rearrange $V_{c}[\rho]$ into a form of exchange interaction~\cite{KS_sma}:
The Coulomb interaction between two charges, 
each having form factor $f^{jk}(y_{0})$ and $g^{mn}(y_{0})$,
is cast into a form of manifest exchange interaction as follows,
\begin{eqnarray}
&&\sum_{\bf p}v^{\rm C}_{\bf p}\! \int\! dy_{0} 
f^{jk}(y_{0}) {:\!\hat{\cal R}^{jk}_{\bf -p}(y_{0}) }\!
\int\! dz_{0}\, g^{mn}(z_{0})\hat{\cal R}_{\bf p}^{mn}(z_{0})\!: ,
\nonumber\\
&=&- {1\over{\bar{\rho}}} \sum_{\bf k}
  \int\! dy_{0}P^{jk;mn}_{y_{0}; {\bf k}} \, 
{:\! \hat{\cal R}^{jn}_{\bf -k}(y_{0}) }  \int\! dz_{0}\, \hat{\cal R}^{mk}_{\bf k}(z_{0})\!:,
\nonumber\\
&\approx& - {1\over{\bar{\rho}}} \sum_{\bf k}
\int\! dy_{0}\,P^{nm;mn}_{y_{0}; {\bf k}}  
:\! \hat{\cal R}^{nn}_{\bf -k}(y_{0})\, \int\! dz_{0}\,  \hat{\cal R}^{mm}_{\bf k}(z_{0})\!:,
\label{D_to_EX}
\end{eqnarray}
where 
$\hat{\cal R}^{mn}_{\bf -p}(y_{0}) 
= \hat{\psi}^{m\dag}(y_{0})\,e^{i{\bf p}\cdot {\bf r}}\hat{\psi}^{n}(y_{0})$
and $\bar{\rho}= 1/(2\pi \ell^2)$. 
The new form factor is written as
\begin{equation}
P^{jk;mn}_{y_{0}; {\bf k}} =\sum_{\bf p}v^{\rm C}_{\bf p} e^{-i\ell^2{\bf p} \times {\bf k}} 
f^{jk}(y_{0}) \, g^{mn} ( y_{0}^{[p_{x}+ k_{x}]}), 
\label{FFcombined}
\end{equation}
where $y_{0}^{[p_{x}+ k_{x}]} \equiv y_{0} - \ell^2 (p_{x} + k_{x})$. 
See Appendix C for a derivation of this formula. 
Relevant to our present analysis is the last line of Eq.~(\ref{D_to_EX}), 
that retains only the leading diagonal charges $\hat{\cal R}^{nn}_{\bf -k}$ 
and $\hat{\cal R}^{mm}_{\bf k}$.

Upon rearrangement, 
the $:\! (\delta_{2} \hat{\rho}_{\bf -p}) \, \delta_{1}\hat{\rho}_{\bf p}\! :$ term
in $\delta_{\rm II} V_{c}$ acquires a form factor
involving  a product
\begin{equation}
-\big[\Xi^{(2)}_{\bf p}e^{i{\bf p\cdot X}} \big]^{nm}\, 
\big[ \Xi^{(1)}_{\bf -p}|_{y_{0} \rightarrow y_{0}^{[p_{x} + k_{x}]}} e^{-i{\bf p\cdot X}} \big]^{mn}.
\label{Xone_Xtwo}
\end{equation}
With $\Xi^{(2)}_{\bf \pm p} \propto \pm p_{y}$, 
this term cancels out the corresponding term 
$\propto -[e^{i{\bf p\cdot X}} \big]^{nm} 
[(\Xi^{(1)}_{\bf -p}\, \Xi^{(2)}_{\bf -p})|_{y_{0} \rightarrow y_{0}^{[p_{x} + k_{x}]}}\, 
e^{-i{\bf p\cdot X}} \big]^{mn}$ 
in $:\! \hat{\rho}_{\bf -p}\, \delta_{2}\delta_{1} \hat{\rho}_{\bf p}\! :$ 
of Eq.~(\ref{dtwo Vc}), apart from terms of $O(C'') \sim O(A''_{0}) + O(v'''_{x}A_{0})$
 beyond our present concern.

The remaining term in $\delta_{\rm II} V_{c}$ involves 
$[S_{2}, \Xi^{(1)}_{\bf p}] \propto  p_{y}v'_{x}(y_{0}) C'(y_{0})$, 
which, upon integration over ${\bf p}$ in $P^{mn;nm}_{y_{0};{\bf k}}$, 
becomes $\propto k_{x}\, O(v'_{x}A'_{0})$. 
(Note here that 
$[e^{i{\bf p\cdot X}}]^{nm}[e^{-i{\bf p\cdot X}}]^{mn} = \gamma_{\bf p}^2\, |f^{mn}_{\bf p}|^2$
is a function of ${\bf p}^2$.)
Such $k_{x}$-dependent terms are sensitive to spatial $x$ variations 
and do not contribute to 
the $x$-averaged (i.e., $k_{x}\rightarrow 0$) current  $j_{x}[y_{0}]$ of our present concern.
For the same reason terms involving odd powers of $p_{y}$ in $\delta_{\rm I} V_{c}$ 
cease to contribute to $j_{x}$.
Here we learn that the Coulomb interaction leaves 
the drift current $j^{\rm (d)}[y_{0}]$ unaffected.
As a result, Eqs.~(\ref{amount_bulkC}),  (\ref{jby_local}) and (\ref{Jd_Ef}) hold as they are~\cite{footnote_Jd}.

On the other hand, the last term ${1\over{2}} v'_{x}(y_{0})\,  ({\bf p} \cdot\! {\bf X})$
in $\Xi^{(1)}_{\bf p}$ survives and leads to the interaction 
\begin{equation}
\Delta V_{c}\approx{1\over{\bar{\rho}}} \sum_{\bf k}Q^{mn}_{\bf k} \!
\int\! dy_{0}\, v'_{x}(y_{0}) \, \hat{\cal R}^{nn}_{\bf -k}(y_{0})
 \int\! dz_{0}\,  \hat{\cal R}^{mm}_{\bf k}(z_{0}),
 \label{bz-Induced}
\end{equation}
with
\begin{eqnarray}
Q^{mn}_{\bf k} &=& {1\over{2}} \sum_{\bf p}
e^{-i\ell^2{\bf p \times k}}\,  v^{\rm C}_{\bf p}\gamma^{2}_{\bf p}\,  w^{mn}_{\bf p},
\label{Qmn}
\\
w^{mn}_{\bf p}&=& -i[({\bf p} \cdot\! {\bf X})\, f_{\bf p}]^{mn}  f^{nm}_{\bf -p}.
\label{w_MN}
\end{eqnarray}
Here the factors $w^{mn}_{\bf p}$ are functions of   $\xi \equiv  {1\over{2}} \ell^2{\bf p}^2$; 
$w_{\bf p}^{mn} = (n!/m!)\, \xi^{m-n}\, \{ \xi L^{m-n+1}_{n}(\xi) - m\, L^{m-n-1}_{n}(\xi)\} L^{m-n}_{n}(\xi)$;  
$w_{\bf p}^{00} = \xi$,  $w_{\bf p}^{10} = w_{\bf p}^{01} = \xi (\xi- 1)$, etc.
Direct calculations yield, e.g.,
\begin{equation}
(Q^{00}_{\bf k\rightarrow 0}, Q^{10}_{\bf 0}, Q^{11}_{\bf 0})
= \tilde{V}_{c}\,({1\over{4}}, {1\over{8}}, {3\over{16}}), \ \ 
 \tilde{V}_{c }= {\alpha_{e}\over{\epsilon_{b}\, \ell}} \sqrt{\pi\over{2}}.
\end{equation}

Obviously this $\Delta V_{c} \propto b_{z}$ represents Coulombic corrections 
to orbital magnetization, and thus also contributes to $j^{\rm (c)}(y)$. 
A simple estimate is to consider the expectation value $\langle \Delta V_{c} \rangle$ 
and approximate the density of a filled level by
$\int dz_{0}\langle  \hat{\cal R}^{mm}_{\bf k}(z_{0})\rangle 
\approx \bar{\rho}\, (2\pi)^2\delta^{2}({\bf k})$, 
i.e., by that in the sample bulk, 
\begin{equation}
\langle \Delta V_{c}\rangle \approx \sum_{n,m} Q^{mn}_{\bf k \rightarrow 0} 
\int\! dy_{0}\,  \ell\, v'_{x}(y_{0}) \, 
\langle \hat{\rho}_{n}(y_{0})  \rangle.
\end{equation}
This leads to the corrections to the magnetic moment 
$m_{n} = -e\ell^2 \omega_{c} (n+ 1/2)$ of an electron in the bulk
\begin{equation}
\Delta m_{n} = e\ell^2 \sum_{m} Q^{mn}_{\bf k \rightarrow 0},
\label{DMn}
\end{equation}
where the sum is  taken over filled levels $\{m\}$.
The Coulomb interaction thus generally works to
reduce orbital magnetization $m_{n}\rightarrow m_{n} + \Delta m_{n}$ 
and the associated current 
$J^{\rm (c)}_{n} \rightarrow -\bar{\rho}\, (m_{n} + \Delta m_{n})$ to some extent.

\section{Graphene}
In this section we consider the case of graphene. 
The electrons in graphene are described by two-component spinors 
on two inequivalent lattice sites. 
They acquire a linear spectrum (with velocity $ v_{\rm F} \sim 10^{6}$m/s) 
near the two inequivalent Fermi points $(K,K')$ in momentum space, 
with an effective Hamiltonian of the form~\cite{Semenoff},  
\begin{eqnarray} 
H &=&
\int d^{2}{\bf x}\, \{ \Psi^{\dag}_{+} {\cal H}_{+}\Psi_{+} + \Psi_{-}^{\dag} {\cal H}_{-}\Psi_{-} \},  
\nonumber\\
{\cal H}_{\pm} &=& 
v_{\rm F}\, (\Pi_{x}\sigma^{1}+ \Pi_{y}\sigma^{2})  \pm \delta m\, \sigma^{3}  - eA_{0},
\label{H_GR}
\end{eqnarray}
where $\Pi_{i}= p_{i}+eA_{i}$ and $\sigma^{i}$ denote Pauli matrices.
The Hamiltonians ${\cal H}_{\pm}$ describe electrons 
in two different valleys $a \in (K,K')$ per spin, and 
$\delta m$ stands for a possible sublattice asymmetry; 
we take $\delta m > 0$, without loss of generality.
Actually, valley asymmetry of a few percent is inferred from experiments~\cite{HuntYY,WBEM} 
using high-mobility graphene/hexagonal boron nitride (hBN) devices.

Let us place graphene in a uniform magnetic field $B_{z}=B>0$ 
and include also weak potentials 
$v({\bf x}) = e \ell\, \{a_{y}({\bf x}) + ia_{x}({\bf x}) \}/\sqrt{2}$ and $A_{0}({\bf x})$.
In the $|\, |n|,y_{0}\rangle$ basis, 
the Hamiltonian ${\cal H}_{+}$ in valley $K$ is written as
\begin{equation}
{\cal H}_{+} = \omega_{c} \left(
\begin{array}{cc}
\mu & -Z \!-i v({\bf x}) \\
-Z^{\dag} \!+iv^{\dag}({\bf x})& -\mu \\
\end{array}\!
\right) - e A_{0}({\bf x}), 
\label{Hv_gr}
\end{equation}
where $Z= (Y+ iP)/\sqrt{2}$; 
\begin{equation}
\omega_{c} \equiv \sqrt{2}\, v_{\rm F}/\ell \ \ {\rm and}\ \ 
\mu \equiv \delta m/\omega_{c}.
\end{equation}

For $v=A_{0}=0$, the electron spectrum forms an infinite tower 
of Landau levels of energy 
\begin{equation} 
\epsilon_{n} =\omega_{c} e_{n} \ {\rm and}\ \ e_{n} \equiv s_{n} \sqrt{|n|+\mu^{2}}
\end{equation}
in each valley (with $s_{n}\equiv  {\rm sgn}[n] = \pm1$), 
labeled by integers $n \in (0,\pm 1, \pm2, \dots)$ and
$y_{0}=\ell^2 p_{x}$, of which only the $n=0$ (zero-mode) levels split in the valley
(hence to be denoted as $n=0_{\mp}$), 
\begin{equation}
\epsilon_{0_{\mp}}= \mp \delta m = \mp \omega_{c}\,  \mu   \ \ {\rm for}\ K/K'.
\end{equation}
Thus, for each integer $|n| \equiv N = 0,1,2, \cdots$ 
(we use capital letters for the absolute values),
there are in general two modes with $n=\pm N$ (of positive/negative energy) 
per valley and spin, apart from the $n=0_{\pm}$ modes.

The eigenmodes in valley $K$ are written as~\cite{KS_LWGS}
\begin{equation}
\Phi^{n}_{y_{0}}({\bf x})|^{K} 
= \big(  \langle {\bf x}  | N\! -\!1, y_{0} \rangle\, b^{n}, \langle {\bf x} | N, y_{0} \rangle\, c^{n} \big)^{\rm t},
\label{psi_n}
\end{equation}
with $(b^{n}, c^{n})^{\rm t}$ given 
by the (normalized) eigenvectors of the reduced (numerical) matrix 
${\cal H}_{+}|_{N}^{\rm red}$
obtained from ${\cal H}_{+}|_{v=A_{0}=0}$ by replacing $Z, Z^{\dag}\rightarrow \sqrt{N}$.
In explicit form,
\begin{equation}
(b^{n}, c^{n}) = {1\over{\sqrt{2}}}\, \left( \sqrt{1+ {\mu\over{e_{n}}} }, -s_{n} \sqrt{1- {\mu\over{e_{n}}} }\right),
\label{bc_muzero}
\end{equation}
and $(b^{0_{-}}, c^{0_{-}})=(0, 1)$.

 One can pass to another valley $K'$ by simply setting $\mu\rightarrow - \mu$ 
 since ${\cal H}_{-} =  {\cal H}_{+}|_{-\mu}$ holds, 
 where  ${\cal O}|_{-\mu}$ signifies reversing the sign of $\mu$ in ${\cal O}$.
 Actually, the electron and hole spectra are intimately related, via electron-hole ($e$-$h$) symmetry, 
 between valleys $(K,K')$ and also within each valley.
 Let $\epsilon^{K}_{n}[\mu,A_{0}]$ denote the spectrum of level $n$ in valley $K$ for a given $(\mu, A_{0})$.
The unitary equivalence $\sigma^{3}\, {\cal H}_{-}\sigma^{3} = -{\cal H}_{+}|_{-A_{0}}$ and
$\sigma^{3}\, {\cal H}_{+}\sigma^{3} = -{\cal H}_{+}|_{-A_{0}, -\mu}$ 
then implies the following relations
\begin{equation}
\epsilon^{K}_{n}[\mu,A_{0}] = -\epsilon^{K'}_{-n}[\mu, -A_{0}] = -\epsilon^{K}_{-n}[-\mu, -A_{0}]
\label{UnitaryEquiv}
\end{equation} 
as well as  $\epsilon^{K'}_{n}[\mu,A_{0}] = \epsilon^{K}_{n}[-\mu, A_{0}]$.    
For notational clarity, we henceforth suppress obvious valley (and spin) labels, 
and mainly present $K$-valley expressions.

Let us now turn on $(v, A_{0})$ and 
expand $\Psi_{+}({\bf x}) =\sum_{y_{0},n} \Phi^{n}_{y_{0}}({\bf x})\, \psi^{n}(y_{0})$ 
in terms of $\{\Phi^{n}_{y_{0}}({\bf x}) \}$.
The one-body Hamiltonian $H_{+}$ is then written as~\cite{KS_LWGS}
\begin{eqnarray}
H &=&\int\! d y_{0}\,  \psi^{m \dag}(y_{0}) {\cal H}^{mn} \psi^{n}(y_{0}),
\nonumber\\
{\cal H} &=& \omega_{c} \big\{ 
-b\, (Z+ iv)\, c - c\, (Z^{\dag} - iv^{\dag})\, b
\nonumber\\
&&+ \mu\,  (b\, b - c\, c) \big\} -  e\, b\, A_{0}\, b- e\, c\,A_{0}\,c, \ \ \ 
\label{H_gr}
\end{eqnarray}
where $v= v({\bf x})$ and $A_{0}= A_{0}({\bf x})$ with ${\bf x} = {\bf X + r}$; 
orbital labels $(m,n)$ now run over all integers $(0, \pm1,\pm2, \cdots)$. 
Here we have introduced condensed notation: 
For ${\cal H}^{mn}$ we interpret, e.g., 
\begin{eqnarray}
b\, Z\, c\,  &\rightarrow&\,   b^{m}\, Z^{M-1,N}\, c^{n}, \ 
b\, v\, c \, \rightarrow\,     b^{m}\, [v({\bf x})]^{M-1,N} c^{n},\nonumber\\
b\, b\,   &\rightarrow&\,  b^{m}\, 1^{M-1,N-1}\, b^{n}, \ 
c\, c\,  \rightarrow\,   c^{m}\, 1^{M,N}\, c^{n},\nonumber\\
c\,A_{0}\,  c\,   &\rightarrow&\,  c^{m}\, [A_{0}({\bf x})]^{M,N}\, c^{n}, {\rm  etc.},
\label{condensed_nt}
\end{eqnarray}
with $M=|m|$,  $N=|n|$, 
 $Z^{MN} \equiv \sqrt{N}\, \delta^{M, N-1}$,
 $[v({\bf x})]^{M,N}\equiv  \langle M|v({\bf x})|N\rangle$, 
 $1^{M,N}\equiv \delta^{M,N}$, etc.

In each $N$ sector, the associated eigenvectors form an orthogonal matrix 
$((b^{N}, c^{N})^{\rm t}, (b^{-N}, c^{-N})^{\rm t})$.
Obviously the row vectors also form an orthonormal set, which we denote 
as $b \sim (b^{N},b^{-N})^{\rm t}$ and $c \sim (c^{N},c^{-N})^{\rm t}$.  
We write their inner products (e.g., $b\cdot b \equiv b^{N}b^{N}+ b^{-N}b^{-N}$) as
\begin{equation}
b  \cdot b  = c \cdot c =1, \ \ b \cdot c = c \cdot b = 0
\end{equation}
for each $N$ and subsequently for all $N=0,1,2, \cdots$. 
[For the $N=0$ sector one only has $c^{0_{\mp}} =\pm 1$ (and $b\, b=0$);
in most cases $b^{0}=0$ is automatically eliminated 
via the associated matrix elements like $c^{m}1^{M,N-1}\, b^{n}$.]
In this way, the orbital space $\{ n\}$ is decomposed into 
two subspaces referring to $(b,c)$.
Note that $(b\, b)^{mn}$ and $(c\, c)^{mn}$, defined in Eq.~(\ref{condensed_nt}), 
act as projection operators,
\begin{eqnarray}
b\,b \cdot b\, b =b\, b,\  c\, c \cdot c\, c = c\, c,\  b\,b+ c\,c = {\bf 1}.
\end{eqnarray}
In addition, $(bb -cc, bc, cb)$ obey formally the same algebra, 
e.g., $[bc, cb] = bb-cc$, as  $(\sigma_{3},\sigma_{+}, \sigma_{-})$; 
$\sigma_{\pm}= (\sigma_{1}\pm i\sigma_{2})/2$.
The SU(2) spinor structure of the Dirac Hamiltonian is thus projected 
onto the $(b,c)$ sectors of infinite dimensions.
Such algebraic features are naturally shared by multilayers of graphene as well.

Inner products play a role in multiplication. Note, e.g., 
$b\, O_{1} b \cdot b\, O_{2} b = b\,  O_{1} O_{2}\, b 
\ (=b^{m} (O_{1} O_{2})^{M-1,N-1}b^{n}$),
$c\, O_{1}c \cdot c\, O_{2}c = c\,  (O_{1} O_{2}) c$  and $O\, b \cdot c\, O' =0$.
These features suggest us how to generalize the $W_{\infty}$ rotations 
$\psi_{U}^{M} = U^{MN}\psi^{N}$ 
in the $|N,y_{0}\rangle$ basis of Sec.~II, to the present spinor case.
We extend them to $\psi^{n}(y_{0})$ by setting 
\begin{equation}
\psi_{U} = {\cal U} \psi\ \   {\rm with}\ \   {\cal U} = b\, U b + c\, U c.
\end{equation} 
This ${\cal U}$, when acting on ${\cal H}$ in Eq.~(\ref{H_gr}), 
induces rotations in the $|N,y_{0}\rangle$ basis, e.g., 
\begin{eqnarray}
{\cal U}\cdot b\, (Z+ iv)\, c \cdot {\cal U}^{\dag} 
&=& b (U(Z + iv)U^{\dag}) c, 
\end{eqnarray}
and ${\cal U}\! \cdot  b\, {\cal O}b\! \cdot {\cal U}^{\dag} =  b\, (U{\cal O}U^{\dag})b$,
while retaining the $(bc, cb,bb,cc)$ outer structures intact, 
\begin{equation}
{\cal U}\, (bc, cb,bb,cc)\,{\cal U}^{\dag} = (bc, cb,bb,cc).
\end{equation}
Let us also project $(Z,Z^{\dag})$ into the $(b,c)$ space by setting
\begin{equation}
{\cal Z} = b\, Z b + c\,  Z c,\ \ {\cal Z}^{\dag} = b\, Z^{\dag} b + c\, Z^{\dag} c,
\end{equation} 
which obey the same algebra as $(Z,Z^{\dag})$, with $[{\cal Z}, {\cal Z} ^{\dag}] =1$; 
${\cal Z}^{mn} \propto \delta^{M,N-1}$, etc.
A given $W_{\infty}$ rotation $U(Z,Z^{\dag})$ in the $(N,y_{0})$ basis
is then immediately promoted to ${\cal U}$ in the $(n, y_{0})$ space by replacing 
$(Z,Z^{\dag})$  $\rightarrow ({\cal Z},{\cal Z}^{\dag})$ in $U$.  
One can thus write the field transformation law as 
\begin{eqnarray}
\psi_{U}^{m}(y_{0})  
&=&  [U({\cal Z}, {\cal Z}^{\dag}; {\bf r})]^{mn}\, \psi^{n}(y_{0}).
\end{eqnarray}
It is also possible to express the Hamiltonian ${\cal H}$ itself 
in terms of $({\cal Z}, {\cal Z}^{\dag})$,
\begin{eqnarray}
{\cal H}[v, A_{0}]
&=& \omega_{c} \big\{\! - bc \cdot ({\cal Z}+ i v) - ({\cal Z}^{\dag}- iv^{\dag}) \cdot cb
\nonumber\\
&&  +  \mu\, (bb - cc) \big\}  -e A_{0},
\label{H-vA-GR}
\end{eqnarray}
where $v$ and $A_{0}$ stand for $v({\bf x})$ and $A_{0}({\bf x})$ 
with obvious replacement
$(Z,Z^{\dag}) \rightarrow ({\cal Z},{\cal Z}^{\dag})$ in the argument ${\bf x}={\bf X +r}$.

The basic framework of $W_{\infty}$ gauge theory, 
including the gauge transformation $G= e^{iS}$, $v^{G}$ and  $A_{0}^{G}$, 
developed in Sec.~II, is now naturally adapted to the present spinor case
by simple replacement $(Z,Z^{\dag}) \rightarrow ({\cal Z},{\cal Z}^{\dag})$.  
In particular, via the gauge ransformation $\psi \rightarrow \psi_{G} =G\, \psi$,
with $G = G[{\cal Z}, {\cal Z}^{\dag}]$,
the Hamiltonian  ${\cal H}[v, A_{0}]$ in Eq.~(\ref{H-vA-GR}) turns into 
${\cal H}^{G} \equiv {\cal H}[v^{G}, A_{0}^{G}]$,
which is {\it diagonal  in the sectors} $\propto \delta^{MN}$ 
to $O(v)$, $O(A_{0})$ and $O(v\partial A_{0})$.
A critical departure from the case of Sec.~II arises when one  eliminates the off-diagonal portion 
$\Delta^{\rm off}{\cal H}^{G}=  eE({\bf r})\,  {\cal Z}^{\dag} + eE^{\dag}({\bf r})\, {\cal Z} + \cdots$ 
by a further transformation $G_{2}= e^{iS_{2}}$, with
\begin{equation}
S_{2}^{mn} = -i {e\over{\omega_{c}}}(e_{m} + e_{n}) 
\{E({\bf r})\,  ({\cal Z}^{\dag})^{mn} - E^{\dag}({\bf r})\, {\cal Z}^{mn} \}.
\end{equation}
This is a sensible unitary transformation but not written as a $W_{\infty}$ transformation. 
[It is the presence of $bc$ and $cb$ that does not allow diagonalization of ${\cal H}[v, A_{0}]$
by $W_{\infty}$ transformations alone.]
This $S_{2}^{mn}$ gives rise to another $O(v\partial A_{0})$ piece, 
such as ${\cal H}^{(2)}_{vA}$ in Eq.~(\ref{H_vA_two}).
For the $N=0$ sector, e.g., it takes the form 
 \begin{eqnarray}
 [{\cal H}^{(2)}_{vA}]^{00} 
 &=&  \mu\,  e{\bf E}({\bf r})\cdot \nabla\,\gamma_{\bf p}\, b_{z}({\bf r}).
 \label{H_vA_twoGR}
 \end{eqnarray}
Such terms contribute to the drift current $j^{\rm (d)}(y)$ 
as integrable and oscillating components,  
which, as discussed in Sec.~III, are sizable only in the vicinities of 
the edge boundaries $y \sim y_{0;n}^{+}$ and 
are made practically invisible by the dominant profile of $j^{\rm (c)}(y)$. 
For this reason and for simplification we omit them in our analysis below.

Let us now denote the diagonal portion as  
$\hat{H} = \int dy_{0}\hat{\psi}^{\dag} \hat{\cal H}\hat{\psi}$, 
with $\hat{\psi} = G_{2}G \psi$ and $\hat{\cal H} = G_{2}{\cal H}^{G}G_{2}^{-1}|^{\rm diag}$, 
and take a static setting 
$v \rightarrow v_{x}(y_{0})$ and $-eA_{0} \rightarrow W(y_{0})$. 
Projecting $\hat{\cal H}$ into diagonal sectors $\{N\}$
then yields
\begin{eqnarray}
\hat{\cal H}^{mn} 
&\stackrel{}{\approx}& \epsilon_{n}  \delta^{mn}  +  \delta^{MN}[k_{N}(\xi)]^{mn} W(y_{0})
\nonumber\\
&&- \delta^{M,N}\, \omega_{c}\, {\textstyle {1\over{2}} }\,  [h^{1}_{N}(\xi)]^{mn} \, b_{z}(y_{0})
\nonumber\\
&&+ \delta^{MN} \ell W'(y_{0})\, [k_{N}(\xi)]^{mn} v_{x}(y_{0}),
\end{eqnarray}
where $[k_{N}(\xi)]^{mn} \equiv b^{m} k_{N-1}(\xi) b^{n} + c^{m } k_{N}(\xi) c^{n}$
and $[h^{1}_{N}(\xi)]^{mn} \equiv (b^{m}\, c^{n} + c^{m}\, b^{n})\, 
(1/\sqrt{N})\,  e^{-{1\over{2}} \xi}\, L_{N-1}^{1}(\xi)$;
$[k_{N}(0)]^{mn} =\delta^{mn}$ and 
$[h^{1}_{N}(0)]^{mn} =\sqrt{N} (b^{m}\, c^{n} + c^{m}\, b^{n})$.

There is a level mixing due to ${\cal H}^{n,-n}$ within the sector $\{N \ge 1\}$, 
but it eventually leads to corrections $\sim O(v'''_{x}W'')$ 
far beyond our present concern.  
The spectra and associated current are therefore read from 
the diagonal components $\hat{\cal H}^{nn} \equiv \hat{\cal H}_{n}$ with $n =\pm N$.

In the $N=0$ sector  $n|^{K}=0_{-}$ (or $n|^{K'}=0_{+}$), 
the associated spectrum and current are read from
\begin{eqnarray}
\hat{\cal H}_{n=0_{\mp}}
&=& \epsilon_{0_{\mp}}[y_{0}]+  \ell W'(y_{0})\, k_{0}(\xi) v_{x}(y_{0}),
\nonumber\\
\epsilon_{0_{\mp}}[y_{0}] &\approx& \mp \omega_{c}\mu  + W(y_{0})
+ {\textstyle {1\over{4}}} \, \ell^2W''(y_{0}). 
\end{eqnarray}
Here we see no single $b_{z} \sim - v'_{x}$ term (since $b^{0}=0$). 
This shows that, in graphene, the zero-mode levels $n = (0_{-}|^{K}, 0_{+}|^{K'})$ 
carry no orbital magnetization and no circulating current $j^{\rm (c)}$ 
while they support a normal amount of drift current 
$j^{\rm (d)} \propto W'(y_{0})$. 
Such features of the $N=0$ sector have been noted and observed~\cite{UKBL} 
in experiment via direct imaging of local currents.  

In the $N \ge 1$ sector  the spectra are written as
\begin{eqnarray}
\epsilon_{n}[y_{0}] 
&=& \epsilon_{n} + [b^{n}b^{n}k_{N-1}(\xi) + c^{n}c^{n}k_{N}(\xi)]W(y_{0}),
\nonumber\\
&\approx& \epsilon_{n}   + W(y_{0}) 
+  {\textstyle {1\over{2}}} (N- {\textstyle {1\over{2}}} \hat{\mu}_{n}) \ell^2 W''(y_{0}),
\end{eqnarray}
where $ \hat{\mu}_{n} \equiv \mu/e_{n} = s_{n} \mu/\sqrt{N + \mu^2}$.
From $\hat{\cal H}_{n}$ one can read off an orbital magnetic moment of an electron, 
$\hat{m}_{n} = e\ell^2\,  \omega_{c}\, {\textstyle {1\over{2}} }\,  [h^{1}_{N}(0)]^{nn}$ 
or
\begin{equation}
\hat{m}_{n} = e\ell^2\,  \omega_{c}\, \sqrt{N} b^{n}c^{n}
= - e \ell^2 \omega_{c} N/(2 e_{n}). 
\end{equation}
This $\hat{m}_{n}$ agrees with one, $-\partial \epsilon_{n}/\partial B$, 
calculated from the spectrum of an electron in the sample interior.

Let us now note Eqs.~(\ref{formula_one}) and (\ref{formula_dLn}).
Then $\hat{\cal H}_{n}$ is readily translated into the edge-current distributions, 
\begin{eqnarray}
\langle j_{n}^{\rm (c)}(y) \rangle &=& -{e\omega_{c}\over{2 \pi}}  {1\over{2 e_{n}}}\!
\int\! dy_{0}\, \partial_{y}\!  \sum_{m=0}^{N-1}  |\phi_{m}(y-y_{0})|^2,
\label{je_GR}
\\
&=&   {e\omega_{c}\over{2 \pi}}{1\over{2 e_{n}}} \sum_{m=0}^{N-1}  |\phi_{m}(y-y^{+}_{0;n})|^2,  
\\
\langle j_{n}^{\rm (d)}(y) \rangle &=& -{e\over{2 \pi}} \int\! dy_{0}\, W'(y_{0})\, R^{nn}(y-y_{0}),
\end{eqnarray}
with the density profile of the $\hat{\psi}^{n}(y_{0})$ mode
\begin{equation}
R^{nn}(y) = b^{n} \, |\phi_{N-1}(y)|^2\, b^{n}+ c^{n} \,|\phi_{N}(y)|^2\, c^{n}.
\end{equation}
The filled domain $\{ y_{0}\le y_{0;n}^{+}\}$ of  each level $n$ is fixed by 
$\epsilon_{n}[y_{0;n}^{+}] =\epsilon_{\rm F}$
for a given $\epsilon_{\rm F}$. 
The diamagnetic circulating current $\langle j_{n}^{\rm (c)}(y) \rangle$ is again explicitly integrated 
to have a profile localized around the edge position $y = y^{+}_{0;n}$. 
For a filled level $n$ it carries the total amount 
\begin{equation} 
J_{n}^{\rm (c)} =   \int dy\, \langle j_{n}^{\rm (c)}(y) \rangle 
=  {e\omega_{c}\over{2 \pi}} \,{N\over{2e_{n}}}= -\bar{\rho}\,\hat{m}_{n}.
\end{equation}
The drift component $\langle j_{n}^{\rm (d)}(y) \rangle$ 
again exhibits a universal growth toward the edge, as in Eq.~(\ref{jby_local}), 
and carries a total amount $J_{n}^{\rm (d)} = \int dy\, \langle j_{n}^{\rm (d)}(y) \rangle$, 
with  
\begin{equation} 
J_{n}^{\rm (d)} =  - {e\over{2 \pi}}\,   W(y_{0;n}^{+})
\approx   - {e\over{2 \pi}}\, (\epsilon_{\rm F} -\epsilon_{n})
\label{jnb_GR}
\end{equation}
for $\epsilon_{\rm F} > \epsilon_{n}$.

There are an infinite number of Landau levels in graphene.
 The neutral $\nu=0$ state, or the "vacuum" state $|0\rangle$, consists of 
 all filled negative-energy levels,
i.e., levels $n$ with $n \le 0_{-}$ in valley $K$ and $n \le -1$ in $K'$.
The spectra and current are to be measured relative to this neutral state (in the bulk).
In this picture, in particular, the empty $n|^{K}=-N$ state (hole) is represented 
as $\hat{\psi}^{-N}(y_{0}) |0\rangle$
and, upon acting on $\hat{H}$, is seen to have the spectrum $- \hat{\cal H}_{-N}$, 
which, according to Eq.~(\ref{UnitaryEquiv}), is equal to
$\hat{\cal H}_{N}|_{-\mu, -A_{0}} = \hat{\cal H}_{N}|_{-A_{0}}^{K'}$.
The $n|^{K}= -N$ hole state therefore has the same spectrum 
as the $n|^{K'}= N$ electron state in $K'$ with the sign of $A_{0}$ reversed. 
It is clear now that the present edge with $W'(y)= -eA'_{0}(y) > 0$ only confines electron levels.
We thus consider only the $\nu>0$ case, with $n=0_{+}|^{K'}, 1,2, \cdots$, below.
The $\nu < 0$ case is simply recovered via $e$-$h$ conjugation 
with $e \leftrightarrow h$ and  $A_{0} \rightarrow -A_{0}$.

For numerical simulations we again use a potential wall of Eq.~(\ref{PotWall})
and adopt $\mu = 0.05$ of valley breaking.
We examine equilibrium currents associated with levels $n=0_{+}|^{K'}$ and $n=(1, 2)|^{K, K'}$, 
with the spin and valley degeneracy $\nu_{n}$ of each level $n$ taken into account, i.e., 
$\nu_{0_{\pm}} =2$ and $\nu_{n} =4$ for $|n|\ge 1$, and with small spin splitting set to zero. 
Let us write $\epsilon_{\rm F} =\omega_{c}\sqrt{n_{\rm f}+\mu^2}$ and 
specify filling of the edge modes by $n_{\rm f}$;
accordingly,   $0< n_{\rm f} <1$ refers to filling of the $n=0_{+}|^{K'}_{\downarrow \uparrow}$
levels near the total filling factor $\nu = 2$, 
$1< n_{\rm f} <2$ to filling of four $n=1$ levels near $\nu = 6$, etc.

 \begin{figure}[bpt]
\begin{center}
\includegraphics[scale=0.64]{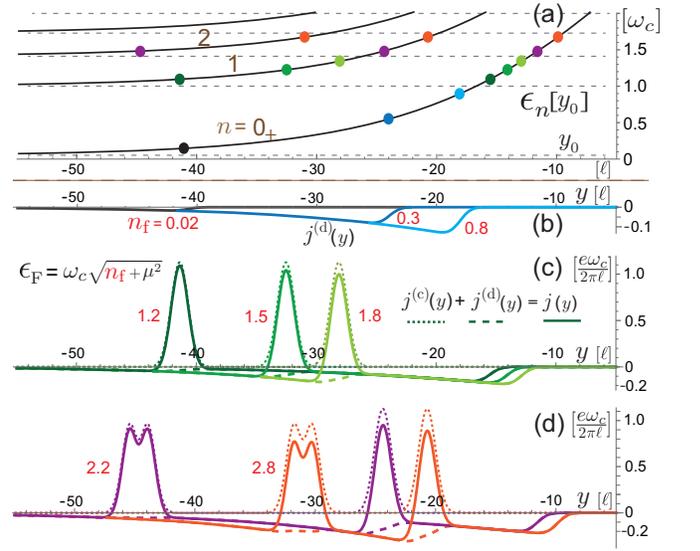}
\end{center}
\vskip-.7cm
\caption{
Graphene.
(a) Level spectra in the edge region. 
(b)~No circulating current $j^{\rm (c)}_{n=0}(y) \rightarrow 0$ arises in the $N=0$ sector.   
(c)-(d)~Edge current distributions,  $j^{\rm (c)}(y) + j^{\rm (d)}(y)$. 
}
\end{figure}

 \begin{figure}[bpt]
\begin{center}
\includegraphics[scale=0.67]{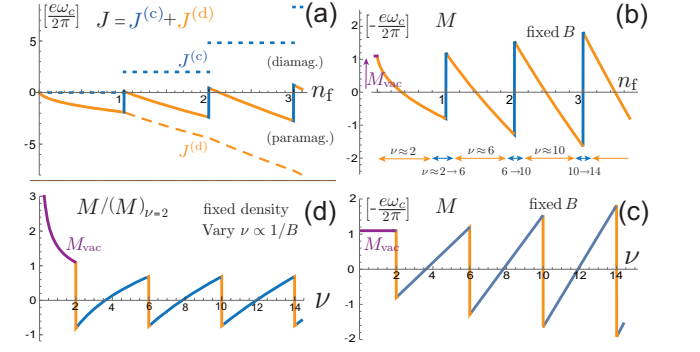}
\end{center}
\vskip-.7cm
\caption{
Currents and magnetization.
(a)~Edge current $J^{\rm (c)}+J^{\rm (d)}$ {\it vs.} edge filling $n_{\rm f}$.
(b)~Orbital magnetization $M = M_{\rm vac} +\langle M^{z}\rangle$ 
[in units of $-(e\omega_{c}/2\pi) < 0$] {\it vs} $n_{\rm f}$.
(c)~Magnetization $M$ plotted as a function of the total filling factor $\nu$.
 (d)~Magnetization {\it vs} $\nu \propto 1/B$ for fixed electron density. 
 }
\end{figure}

Figure~3 presents the current distributions  $j^{\rm (c)}(y)$ and $j^{\rm (d)}(y)$ 
associated with levels $n =0_{+}, 1, 2$. The way the current distributions [in 3(b) - 3(d)]
change with increasing filling $\sim \epsilon_{\rm F} \sim n_{\rm f}$ 
is roughly the same as in the case of Fig.~2,
and is again in clear one-to-one correspondence with
the edge spectra [in 3(a)]. 
A clear difference is the absence of circulating current 
$j^{\rm (c)}_{n=0} \rightarrow 0$ in the $N=0$ sector.
The decrease of drift current $j^{\rm (d)}_{n=0}(y)$ in the vicinity of 
edge boundary $y \sim y_{0;0}^{+}$ is now visible 
but the effect of the integrable component in Eq.~(\ref{H_vA_twoGR}) 
is too weak $\sim O(\mu)$ to be noticeable in the figure. 
Some other differences lie in level-specific profiles of $j^{\rm (c)}_{n}(y)$ 
and their slower growth $\propto \sqrt{N}$ with $N$.

In graphene the paramagnetic drift component $j^{\rm (d)}$ dominates over  $j^{\rm (c)}$
in the total amount, as shown in Fig.~4(a).
This does not mean that graphene exhibits orbital paramagnetism. 
A key fact is that the $\nu=0$ vacuum state, consisting of filled negative-energy sea, 
has an intrinsic quantum response~\cite{ks_emResp}.
In a magnetic field the filled negative-energy sea has the energy density
\begin{equation} 
\epsilon^{B}_{\rm vac} 
=  \bar{\rho}\, \{ -\sum_{n=1}^{N_{\rm cut}}\nu_{n} \epsilon_{n}- \nu_{0_{-}}|\delta m| \}. 
\end{equation}
This is to be compared with the $B\rightarrow 0$ vacuum energy 
$\epsilon^{B=0}_{\rm vac} = -2 v_{\rm F} \sum_{\bf k}\sqrt{{\bf k}^2 + (\delta m)^2}$,
with the Fermi momentum $k_{\rm F}$ chosen to give the same number of negative-energy states,
$N_{\rm s} = k_{\rm F}^2/(2\pi)= (2N_{\rm cut}+1) \bar{\rho}$. 
The deviation $\epsilon_{\rm vac} =\epsilon^{B}_{\rm vac} -\epsilon^{B=0}_{\rm vac}$ 
is finite for $N_{\rm cut}\rightarrow \infty$ and is an observable energy shift~\cite{ks_emResp},  
\begin{equation} 
\epsilon_{\rm vac} = \bar{\rho}\, \omega_{c} \left\{ - 4\, \zeta(-1/2) -2 |\mu|  + O(\mu^2)\right\} >0,
\end{equation}
with a zeta function $-\zeta(-1/2) = \zeta(3/2)/(4\pi) \approx 0.2079$.
This expression for $\epsilon_{\rm vac} \propto B^{3/2}$ 
was also encountered earlier in thermodynamic calculations~\cite{McC,SGB,GGC}.

Thus in graphene the $\nu=0$ vacuum state  has an intrinsic diamagnetic response 
with no associated current (or, with no charge carriers), and leads to 
the magnetization per area 
$M_{\rm vac} =-\partial \epsilon_{\rm vac}/\partial B \propto -\sqrt{B}$,
with
\begin{equation} 
M_{\rm vac} = - {e\omega_{c}\over{2\pi}} \left\{- 6\, \zeta(-1/2) -3 |\mu|  + O(\mu^2)\right\} < 0;
\end{equation}
$- 6\zeta(-1/2) \approx 1.2474$. 
With this vacuum contribution included, 
the magnetization $M = M_{\rm vac} +\langle M^{z}\rangle$ 
with $\langle M^{z}\rangle = -J^{\rm (c)} -J^{\rm (d)}$ oscillates
between diamagnetism and paramagnetism (the de Haas-van Alphen oscillations)
with increasing filling $\nu$ under fixed $B$, as depicted in Figs.~4(b) and 4(c).
Note also Fig.~4(d) which shows that, when one approaches the $\nu=0$ state by increasing $B$ 
under fixed electron density, 
$M$ increases rapidly due to $M_{\rm vac}$, as observed recently in experiment~\cite{BWFP}.

Let  us finally examine the effects of Coulomb exchange interactions.
The charge density $\rho_{\bf -p} \propto e^{i{\bf p}\cdot {\bf x}}$ is a  function of ${\bf x}$ 
and is readily promoted to the $\{n,y_{0};b,c\}$ space 
by replacement $(Z,Z^{\dag}) \rightarrow ({\cal Z}, {\cal Z}^{\dag})$ in ${\bf x}$.
The analysis developed in Sec.~IV applies to the present case of graphene equally well.
Actually, some extra care is needed to handle 
$\Xi^{(2)}_{\bf p} = [S_{2}, e^{i{\bf p \cdot x}}]e^{-i{\bf p \cdot x}}$,
which, unlike one in Eq.~(\ref{Xi-two}), acquires a matrix structure 
$[\Xi^{(2)}_{\bf p}]^{mn} \not= (\cdots)\, \delta^{mn}$. 
In Sec.~IV we have retained terms up to $O(v'A'_{0})$ for $\delta_{\rm II}V_{c}$.
In reality it suffices to keep terms to $O(vA'_{0})$ to determine $O(\tilde{V}_{c})$ corrections 
to the integrated drift current $J^{\rm (d)}$. 
Let us adopt this simplification and set $\Xi^{(1)} \rightarrow p_{y}v_{x}(y_{0})$ 
in handling $\delta_{\rm II}V_{c}$.
Then the discussion presented around Eq.~(\ref{Xone_Xtwo}) goes through 
and the  $O(vA'_{0})$ terms combine to vanish, 
irrespective of the form of $\Xi^{(2)}_{\bf p} \sim O(A'_{0})$.

One thus eventually reaches the same conclusion as before:  
(i)~The amount of drift current $J^{\rm (d)}$ remains unaffected by the Coulomb interaction. 
(ii)~The circulating current $J^{\rm (c)}$ is affected, 
and the many-body corrections are again cast in the form $\Delta  V_{c}$ of Eq.~(\ref{bz-Induced}), 
with $w^{MN}_{\bf p}$ in Eq.~(\ref{w_MN}) replaced by
\begin{equation}
w^{mn}_{\bf p}
= -i [( p {\cal Z}^{\dag} + p^{\dag} {\cal Z})\,  g_{\bf p}]^{mn}g_{\bf -p}^{nm},
\label{w_mn}
\end{equation}
where $g_{\bf -p}=  b \, f_{\bf -p} b + c\, f_{\bf -p} c$.
The corrections again take the form of Coulombic orbital magnetization induced by filled levels, 
though, now by an infinite number of levels in the Dirac sea. 
Still it is possible to show by direct calculations that the $\nu=0$ vacuum state 
acquires no such Coulombic corrections; see Appendix~D.

In consequence, when the $n=0_{+}|^{K'}$ level 
(with spin degeneracy $\nu_{0_{+}} =2$) is filled, i.e., 
as  $\nu \rightarrow 2$, 
the electrons in the $n=0_{+}$ level will feel the same amount 
$\Delta m_{0_{+}}= (1/4) e\ell^2 \tilde{V}_{c}$ (per spin) 
of many-body correction to magnetization $M$ as in the case of Sec.~IV. 
Accordingly, in graphene, when the lowest $N=0$ level is filled, 
a weak circulating current $j^{\rm (c)}_{0_{+}}$ of many-body origin will arise 
around $y\sim y_{0;0}^{+}$
and flow in the same direction as the drift current $j^{\rm (d)}_{0_{+}}$.

\section{Summary and discussion}

In equilibrium, QH electron systems support two species of current,
$j^{({\rm c})}$ and $j^{({\rm d})}$,
forming an alternating pattern of counterflowing channels of current
along the sample edges, 
as predicted earlier theoretically~\cite{GV}
and as observed recently in experiment~\cite{UKBL} 
by use of a nanoscale magnetometer.
In this paper, inspired by such early and recent works, 
we have examined distinctive features of these edge currents 
and derived their real-space distributions.

The drift current $j^{({\rm d})}(y)$ is essentially a Hall current driven 
by a local edge field $\propto W'(y)$.
Its total amount $J^{\rm (d)}$ is universally fixed 
by a sum of Hall potentials $\propto \sum_{n}(\epsilon_{\rm F} -\epsilon_{n})\, \nu_{n}$ 
across the edge region and is left unaffected by the Coulomb interaction, 
as in the QH effect.

Associated with cyclotron motion of an electron is a microscopic diamagnetic current.
This current cancels out locally in a densely populated domain 
while it leaves uniform magnetization inside and 
a circulating current $j^{\rm (c)}(y)$ along its periphery. 
The narrow profile $\Lambda_{n}(y-y_{0;n}^{+})$ of $j_{n}^{\rm (c)}(y)$ 
and the integrated amount $J_{n}=-\bar{\rho}\,\hat{m}_{n}$ are universally fixed 
by the level index $n$ and $B$, reflecting the underlying quantized cyclotron motion, 
although the Coulomb interaction affects them to some extent. 
Intriguingly, as noted in Sec.~V,  
in graphene the lowest Landau level $\ni n=0_{\mp}$ supports no orbital magnetization 
and hence no circulating current $j_{n=0}^{({\rm c})}(y) \rightarrow 0$ at the one-body level 
while a weak current of many-body origin will arise and flow in the same direction as 
 the drift current $j^{\rm (d)}_{0_{+}}(y)$.   
 It will be a challenge to detect such direct signals of interaction in graphene.

Observation of the orbital magnetization $M$ offers 
an indirect way of detecting the equilibrium currents.  
In particular, its paramagnetic portion of response $M^{\rm (d)}$ 
around integer fillings is due to the edge-driven drift current $J^{\rm (d)}$,
and thus implies the presence of the edge states,
which are invisible~\cite{Peierls} in thermodynamic calculations.

Crucial to our analysis is the use of a refined description of QH systems 
as a $W_{\infty}$ gauge theory, which allows one to handle diagonalization of 
the many-body Hamiltonian according to the resolutions of external probes 
and in a manifestly gauge invariant way.
One can thereby define, e.g., the charge, current and  magnetization densities 
in the form of diagonal operators.  Such a framework will also find applications 
in some nonperturbative treatments 
(such as the Hartree-Fock and single-mode approximations) 
as well as in perturbation theory.

\acknowledgments
 This work was supported in part by JSPS KAKENHI Grant Number JP21K03534.

\appendix

\section{$W_{\infty}$ gauge transformation $G$}

In this Appendix, we outline the derivation 
of the $W_{\infty}$ gauge transformation $G = e^{iS}$ in Eq.~(\ref{Gfirst}).
Let us first divide the $W_{\infty}$ generators $\{ (Z^{\dag})^{r} Z^{s}\}$
into three groups (of diagonal/off-diagonal matrices): 
(i)~${\cal F}_{s}=(Z^{\dag})^{s} Z^{s}$,  (ii)~$(Z^{\dag})^{r}{\cal F}_{s}$, (iii)~${\cal F}_{s}Z^{r}$
for integers $s\ge 0$ and $r\ge 1$.
 We expand $S$ of $G = e^{i S}$ in the form
 \begin{equation}
 S = \sum_{s=1}^{\infty} a_{s}{\cal F}_{s} 
 +\sum_{s=0}^{\infty}\sum_{r=1}^{\infty} 
 \{ b_{sr} (Z^{\dag})^{r}{\cal F}_{s} + b_{sr}^{*} {\cal F}_{s}\, Z^{r} \},
 \end{equation}
 where  $a_{s}$ are real coefficients and $b_{sr}$ are complex ones.
 The gauge field $v({\bf x})$  is expanded in multipoles as 
\begin{eqnarray}
v({\bf x}) 
&=& \sum_{s=0} \Big[ {1\over{(s!)^2}}{\cal F}_{s} + {\cal D}_{s}[\partial] \, \Big]\,
 \gamma_{\bf p}(\partial^{\dag} \partial)^{s}  v({\bf r}),
\label{vx_expand} \\
{\cal D}_{s}[\partial] &=&  \sum_{r=1} {1\over{s!\, (s+r)!}}
\{ (Z^{\dag})^{r} {\cal F}_{s} \partial^{r}+ {\cal F}_{s} Z^{r} (\partial^{\dag})^{r} \}. \ \ 
\end{eqnarray}

Evaluating the commutator $[Z,S]$ for the transformed field 
$v^{G} \approx  v - [Z,S]$ to $O(v)$
yields 
\begin{eqnarray}
[Z,S] &=& (s+1)  a_{s+1}{\cal F}_{s}Z 
\nonumber\\
&&+ (s+r) b_{sr} (Z^{\dag})^{r-1}{\cal F}_{s} + s\, b_{sr}^{*} {\cal F}_{s-1}\, Z^{r+1},
\nonumber\\
&=& (s+1) b_{s1} {\cal F}_{s} + (s+r+1) b_{s, r+1} (Z^{\dag})^{r} {\cal F}_{s} 
\nonumber\\
&& + (s+1)a_{s+1}{\cal F}_{s}\, Z +(s+1)\, b_{s+1, r}^{*} {\cal F}_{s}\, Z^{r+1},\ \  \ \ \ 
\label{I_com}
\end{eqnarray}
where, in each line, summation  
is made over repeated integers $s\ge 0$ and $r\ge 1$ (with the sign $\sum_{s,r}$ suppressed).

Let us now compare Eqs.~(\ref{vx_expand}) and~(\ref{I_com}). 
On choosing 
\begin{equation}
 b_{sr}  = {1\over{s!\, (s+r)!}}\,  \gamma_{\bf p}  \partial^{r-1}(\partial^{\dag} \partial)^{s}v({\bf r})
\end{equation}
for  $s\ge 0$ and $r\ge 1$, 
one can remove the ${\cal F}_{s}$ and $(Z^{\dag})^{r} {\cal F}_{s}$ terms from $v^{G}$. 
With this choice of $b_{sr}$, the ${\cal F}_{s}Z^{r}(\partial^{\dag})^{r}$ terms ($r\ge 1$) in $v^{G}$ 
take a form proportional to 
$(\partial^{\dag})^{r-1} (\partial^{\dag} v- \partial v^{\dag}) = -i (\partial^{\dag})^{r-1} b_{z}({\bf r})$.
As for the remaining ${\cal F}_{s}Z\partial^{\dag}$ terms, 
the choice of real parameters $a_{n}$, 
\begin{equation}
(s+1)\, a_{s+1}= {1\over{s!\, (s+1)!}} \gamma_{\bf p} (\partial^{\dag} \partial)^{s}
{\rm Re}[\partial^{\dag} v({\bf r})]
\end{equation}
for $s\ge 0$, eliminates the unfavored real part 
${\rm Re}[\partial^{\dag} v({\bf r})] ={1\over{2}} \nabla\! \cdot\! v({\bf r})$
from the coefficient $\partial^{\dag} v({\bf r})$, 
leaving terms $\propto -i {1\over{2}}b_{z}({\bf r}){\cal F}_{s}Z$.
Some adjustment of notation then leads to 
the  expressions for $S$ in Eq.~(\ref{Gfirst}) and $v^{G}$ in Eq.~(\ref{vG}).

\section{Some formulas} 

In this appendix we present the derivation of Eqs.~(\ref{formula_one}) - (\ref{formula_R}). 
The Fourier transform of $k_{n}({1\over{2}}q^2)$, 
\begin{equation}
\sum_{q}e^{-i q y}k_{n}(\xi)= \sum_{q}e^{-i q y - {1\over{4}} q^2} L_{n}(\xi) = |\phi_{n}(y)|^2,
\end{equation}
with $\xi = {1\over{2}}q^2$, is readily verified. 
Let us note the relation $h_{n}(\xi) = -\partial_{\xi} k_{n}(\xi)$ 
and set $\Lambda_{n}(y) = \sum_{q}e^{-i q y}h_{n}(\xi)$.
One can write $\partial_{q}k_{n}(\xi) =q\,\partial_{\xi}k_{n}(\xi) = -q\, h_{n}(\xi)$, 
which implies the relation
\begin{equation}
\partial_{y}\Lambda_{n}(y) = -i\sum_{q}e^{-iqy}q\, h_{n}(\xi) = -y |\phi_{n}(y)|^2.
\end{equation}
This immediately leads to the Fourier transform of
$k_{n}(\xi) - q^2  h_{n}(\xi)$ in Eq.~(\ref{formula_R}).

On the other hand, examining the action of $Z$ and $Z^{\dag}$ on $\phi_{n}(y)$
reveals a formula
\begin{equation}
y\phi_{n}^2 = -\partial_{y}\{ {\textstyle {1\over{2 }}} \, \phi_{n}^2 +  \phi_{n-1}^2 + \cdots +  \phi_{0}^2\},
\end{equation}
with $\phi_{n} = \phi_{n}(y)$ for short. 
One can now identify the expression for $\Lambda_{n}(y)$ in Eq.~(\ref{formula_one}) 
and fix the Fourier transform of 
$e^{-{1\over{2}} \xi} L_{n-1}^{1}(\xi)= h_{n}(\xi) - {1\over{2}} k_{n}(\xi)$ 
in Eq.~(\ref{formula_dLn}).

\section{Field rearrangement} 

In this appendix we outline the derivation of Eq.~(\ref{D_to_EX}) for field rearrangement. 
Note first that, in the $|y_{0} \rangle$ basis, the plane wave $e^{-i{\bf p \cdot r}}$ is a unitary matrix, 
with elements
\begin{equation}
\langle y_{0}| e^{-i{\bf p \cdot r}}|y'_{0}\rangle 
= \delta (y_{0} - y'_{0} + \ell^{2}p_{x})\, 
e^{-i{1\over{2}}p_{y} (y_{0} + y'_{0} )}.
\end{equation}
They obey the completeness relation
\begin{equation}
\sum_{\bf p}\,  \langle y'_{0}| e^{-i{\bf p \cdot r}}|y_{0}\rangle
\langle z_{0}| e^{i{\bf p \cdot r}}|z'_{0}\rangle 
= \bar{\rho}\, \delta (y_{0}-z_{0})\,  \delta (y'_{0}-z'_{0}),
\end{equation}
as verified directly, 
where 
$\bar{\rho} \! \equiv 1/(2\pi \ell^2)$.
This relation allows one to express the field product $\psi^{m\dag} \psi^{n}$
in terms of charge operators
$R^{mn}_{\bf -p} =\int dy_{0}\,  \psi^{m\dag}e^{i{\bf p} \cdot {\bf r}} \psi^{n} = \sum_{y_{0}, y'_{0}}
 \psi^{m\dag}(y_{0}) \langle y_{0}| e^{i{\bf p} \cdot {\bf r}}|y'_{0}\rangle \psi^{n}(y'_{0})$,
\begin{equation}
\bar{\rho}\, \psi^{m\dag}(y_{0}) \psi^{n} (y'_{0}) 
=  \sum_{\bf p}\,  \langle y'_{0}| e^{i{\bf p \cdot r}}|y_{0}\rangle\, R^{mn}_{\bf p}.
\label{inversionformula}
\end{equation}

Let us  substitute the above formula into 
the $\hat{\psi}^{m \dag}(z_{0}) \hat{\psi}^{k}(y'_{0})\, g^{mn}(z_{0})$ portion 
of the first equation in Eq.~(\ref{D_to_EX}).  This yields 
the following expression between $\hat{\psi}^{j\dag}(y_{0})$ and $\hat{\psi}^{n}(z'_{0})$,
\begin{equation}
\sum_{\bf k} \langle y_{0}| e^{i{\bf p\cdot r}}  e^{i{\bf k\cdot r}} g^{mn}(y_{0})
e^{-i{\bf p\cdot r}}|z'_{0}\rangle\, \hat{R}^{mk}_{\bf k}. 
\end{equation}
Reducing it by making use of relations 
$e^{i{\bf p\cdot r}}  e^{i{\bf k\cdot r}} 
=  e^{-i\ell^2{\bf p} \times {\bf k}}\, e^{i{\bf k\cdot r}}  e^{i{\bf p\cdot r}}$
and $e^{i{\bf p\cdot r}}y_{0} e^{-i{\bf p\cdot r}}= y_{0}- \ell^2 p_{x}$ 
then leads to  the key formula in Eq.~(\ref{D_to_EX}).

\section{Coulombic corrections in graphene}
In this appendix, we discuss the absence of Coulombic corrections to orbital magnetization 
in the $\nu=0$ ground state in graphene.  
Note first that $w^{mn}_{\bf p}= \Omega_{\bf p}^{mn} g_{\bf -p}^{nm}$ 
in Eq.~(\ref{w_mn}),
with  $\Omega_{\bf p}^{mn}= -i  [( p {\cal Z}^{\dag} + p^{\dag} {\cal Z})\,  g_{\bf p}]^{mn}$,
 satisfy the relation
$\sum_{n={\rm all}}w^{mn}_{\bf p}= -i (p{\cal Z}^{\dag} + p^{\dag}{\cal Z})^{mm}=0$;
this means that,  when all levels are filled, the $O(\tilde{V}_{c})$ corrections to orbital magnetization 
combine to vanish for each level $m$.
Let us next rearrange the sum $\sum_{n\le -1}$ in the form 
${1\over{2}}[ \sum_{n={\rm all}} -\sum_{n=0} - D]$ with
$D = \sum_{n\ge 1} - \sum_{n\le -1}$. This yields
\begin{eqnarray}
\kappa_{\bf p}^{m} &\equiv& \sum_{n\le -1}w^{mn}_{\bf p} 
= - {1\over{2}} \Big( \Omega_{\bf p}^{m0}g_{\bf -p}^{0m} + D^{m}_{\bf p} \Big),
\nonumber\\
D^{m}_{\bf p}  &=& \sum_{n \ge 1} (\Omega_{\bf p}^{mn}g_{\bf -p}^{nm} 
- \Omega_{\bf p}^{m, -n}g_{\bf -p}^{-n, m} ).
\end{eqnarray}
In view of $e$-$h$ conjugation which interchanges $K \leftrightarrow K'$, 
one readily sees that  ${\cal Z}^{mn}|^{K} = {\cal Z}^{-m, -n}|^{K'}$, 
$g^{mn}_{\bf p}|^{K'} = g^{-m, -n}_{\bf p}|^{K}$, etc.,
which then imply  $D^{-m}_{\bf p}|^{K} = - D^{m}_{\bf p}|^{K'}$.

For the $0_{-}|^{K}$ and $0_{+}|^{K'}$ levels, in particular, 
one finds $D^{0_{-}}_{\bf p}|^{K} = -D^{0_{+}}_{\bf p}|^{K'}$
and $\Omega_{\bf p}^{00}g_{\bf -p}^{00} = w_{\bf p}^{00}={1\over{2}}\,\ell^2{\bf p}^2$,
so that 
\begin{equation}
\kappa_{\bf p}^{0_{\mp}} = {1\over{2}} ( -w^{00}_{\bf p} \mp D^{0_{-}}_{\bf p}). 
\end{equation}
This shows that the {\it empty} $N=0$ sectors $\ni (0_{-}|^{K}, 0_{+}|^{K'})$ 
feel an orbital magnetic moment $\propto - w^{00}_{\bf p}$ 
coming from the filled valence band (with $n\le -1$).
Thus, when the $0_{-}|^{K}$ level is filled, 
an extra moment $\propto w^{00}_{\bf p}$ is added  
and the resulting $\nu=0$ "vacuum" state has no $O(\tilde{V}_{c})$ correction 
to orbital magnetization, as expected.

\vskip -0.4cm



\begin{thebibliography}{99}
\bibitem{footnote_Jd}
Note, in this connection, that the level spectra $\epsilon_{n}(y_{0})$ in general 
acquire $O(\tilde{V}_{c})$ self-energy corrections $\delta\epsilon_{n}$ 
so that the edge boundaries $y_{0;n}^{+} = y_{0;n}^{+}(\epsilon_{\rm F})$ 
are modified accordingly:
still, these equations, as functions of $y_{0;n}^{+}$ or $\epsilon_{\rm F}$, remain as they are.
\end{thebibliography}
\end{document}